\pdfoutput=1
\documentclass[conference]{IEEEtran}
\IEEEoverridecommandlockouts
\usepackage{cite}
\usepackage{amsmath,amssymb,amsfonts}
\usepackage{graphicx}
\usepackage{textcomp}
\usepackage{xcolor}
\usepackage[hyphens]{url}
\usepackage{array}
\usepackage[caption=false,font=footnotesize,labelfont=sf,textfont=sf]{subfig}
\usepackage{stfloats}
\usepackage{url}
\usepackage{verbatim}
\usepackage{bm}
\usepackage{tikz}
\usepackage{pgfplots}
\usepackage{pgfplotstable}
\pgfplotsset{compat=1.18}
\usepgfplotslibrary{statistics}
\usepgfplotslibrary{groupplots}  
\usepackage{tikzviolinplots}
\usepackage{tabularx}
\usepackage{booktabs}
\usepackage{booktabs}       
\usepackage{csvsimple}      
\usepackage{array}          
\usepackage{graphicx}       
\usepackage{xcolor}         
\usepackage{colortbl}
\usetikzlibrary{patterns}
\usepackage{hyperref}
\usepackage{multirow}

\usepackage{longtable}

\usepackage{listings}

\def\BibTeX{{\rm B\kern-.05em{\sc i\kern-.025em b}\kern-.08em
    T\kern-.1667em\lower.7ex\hbox{E}\kern-.125emX}}
\begin{document}

\pdfpagewidth=8.5in
\pdfpageheight=11in

\newcommand{\iscasubmissionnumber}{311}

\pagenumbering{arabic}

\title{\textsc{CacheForge}: 
LLM-Guided End-to-End Generative Cache Replacement Policy for Performance and 
Hardware Efficiency
}


\author{
Kaushal Mhapsekar, Bita Aslrousta, Brijesh Kumar Bhayana,
Paula Contreras, \\ 
Azam Ghanbari, Ethan Goodman, 
Anna Andriiko, Samira Mirbagher Ajorpaz \\
Department of Electrical and Computer Engineering\\
North Carolina State University, Raleigh, NC, USA\\
\{ kmhapse, baslrou, bkbhayan, pcontre, aghanba2, ecgoodm2, aandrii, smirbag\}@ncsu.edu
}

\maketitle
\thispagestyle{plain}
\pagestyle{plain}

\begin{abstract}
Modern cache replacement designs saturate because they operate within fixed representational structures
, hand-crafted and heuristic based feature-engineered predictors, or offline imitation models that cannot generate new decision logic on their own. At the same time, replacement is shaped by the causal interaction of prefetching, thrashing, spatial locality, and access-type behavior, producing an enormous design space that is difficult to traverse manually. Prior approaches typically rely on heuristics, parameter tuning, or imitation of an offline optimal policy, capturing correlations rather than synthesizing new mechanisms. As a result, their performance gains often plateau and they overfit under dynamic workload conditions.

\textsc{CacheForge} is the first framework to \emph{evolve} cache-replacement policies end-to-end by embedding a large language model inside a governed hardware-aware loop. In each iteration, the LLM proposes new C++ replacement logic, the policy is evaluated under a trace-based CRC-2 ChampSim simulator, and the framework enforces feasibility through reward shaping, structural checks, dynamic mutation, temperature scheduling, and cross-policy crossover. This closed-loop \textit{generation-evolution loop} specifically designed for cache replacement policy enables the discovery of compact policies that satisfy hardware constraints while exploring algorithmic transformations beyond fixed predictor structures.

Across SPEC CPU2006, \textsc{CacheForge} outperforms all CRC-2 baselines. It improves the total hit rate by \textbf{27.36\%, 19.69\%, 13.72\%, 13.15\%, 11.83\%, and 5.73\%} over MPPPB, ReD, Hawkeye, SHiP++, LIME, and LRU, respectively. On memory-intensive workloads, it increases IPC by \textbf{10.15\%, 7.89\%, 6.34\%, 3.64\%, 3.12\%, and 2.71\%} over LRU, MPPPB, LIME, ReD, SHiP++, and Hawkeye. 

\end{abstract}

\section{Introduction}

Off-chip DRAM accesses cost hundreds of cycles; consequently, a small number of exposed LLC misses can determine end-to-end performance. Replacement policies therefore remain a critical optimization point: recent work shows that carefully modifying eviction decisions can still yield double-digit IPC gains. For example, Mockingjay~\cite{mockingjay} improves IPC despite worsening MPKI by suppressing prefetch pollution in specific sets, while PACIPV~\cite{pacipv} improves IPC by $1.1$--$3.3\%$ by retaining lines whose misses would directly stall the core. These results demonstrate that the effectiveness of a replacement policy depends on \emph{which} misses are exposed to the critical path, not on aggregate hit rate.

Modern processors amplify this effect. Every LLC access interacts with speculation, coherence, prefetching, DRAM scheduling, and instruction-stream behavior~\cite{pacipv,mockingjay,hermes}. Evicting a line that feeds a dependence chain can introduce a long stall even if total hit rate improves; retaining a line that supports forward progress can eliminate a stall even if MPKI increases elsewhere. Replacement is therefore a causal decision problem influenced by dynamic interactions, rather than a static recency heuristic.

Despite this \textit{causal} complexity, prior automated designs operate within human-defined representational bounds. The IPV framework~\cite{ipv} searches insertion and promotion vectors, a space of $16^{17} \approx 3\times 10^{20}$ possibilities. Multiperspective Placement Promotion and Bypass (MPPPB~\cite{mpppb}) explores over $10^{110}$ combinations across sixteen expert-curated feature families, requiring nearly ten CPU-years of search. PACIPV~\cite{pacipv} restricts exploration to $1024$ RRIP-derived vectors to make exhaustive search tractable. These efforts demonstrate both the promise and saturation of fixed design spaces: no matter how large the parameter space is, the algorithmic structure is fixed in advance.

Replacement policy design has therefore converged on two classes of approaches.  
First, manually engineered heuristics such as DIP~\cite{dip}, RRIP~\cite{rrip}, SHiP~\cite{ship}, and ReD~\cite{red} encode expert insight into compact predictors, but cannot alter their control flow or metadata structures.  
Second, correlation-driven learning frameworks such as Hawkeye~\cite{hawkeye}, Glider~\cite{glider}, Mockingjay\cite{mockingjay} and PARROT~\cite{parrot},  imitate desirable behavior from traces, but their eviction logic is frozen and cannot adapt based on new causal feedback.  
Across both categories, the fundamental limitation is the same: the \emph{algorithmic structure} of the policy is fixed, and search is confined to parameters within that structure.

This limitation arises because replacement is simultaneously a pattern-recognition problem (predicting reuse, phase changes, and prefetch usefulness) and a causal decision-making problem (determining which evictions alter future exposed misses). Prior approaches strengthen one side of this decomposition, pattern recognition or decision heuristics, but none can evolve both. Even IPV and MPPPB, which perform causal evaluation in simulation, are confined to rigid representations whose algorithmic frames cannot change.

This motivates a natural question: what if a replacement policy could evolve \emph{its own} algorithmic structure, its decision flow, metadata rules, and control logic, while simultaneously adapting its finer-grain behaviors to the patterns it observes? What if policies were not tuned, but \emph{grown}?

We introduce \textsc{CacheForge}, a framework that answers this question by treating a replacement policy as \emph{mutable code}. A large language model generates candidate C++ mechanisms whose logic may differ structurally from their predecessors; a trace-based simulator provides causal performance feedback; and a governed evolutionary loop enforces hardware constraints and prevents unbounded growth. Unlike prior work\cite{mpppb}, \textsc{CacheForge} is not restricted to a fixed predictor template: both algorithmic structure and pattern-interpreting logic can evolve.


Our experiments show that a naive LLM cannot accomplish this task on its own. The solution is far from obvious: without explicit governance, the model consistently fails in ways that render autonomous evolution infeasible. It hallucinates code, emits logic that does not compile or violates interface contracts, and frequently produces mechanisms that explode in metadata size or ignore feasibility constraints altogether. When seeded with existing policies, it tends to imitate them too literally and loses exploratory novelty; when unseeded, it repeats the same structural mistakes across generations. Without a structured memory hierarchy, it oscillates or reverts to previously invalid patterns; with naive memory, it collapses into a single design lineage and becomes trapped in local minima. Without carefully shaped rewards, it aggressively maximizes hit rate, occasionally at the cost of IPC by protecting prefetch or writeback lines. Without objective constraints, it invents simplistic but unbounded mechanisms that violate hardware budgets. In short, an unguided LLM either diverges, overfits, collapses, or produces policies that are architecturally meaningless, demonstrating that a governed, hardware-aware evolutionary loop is not optional, but necessary for reliable discovery.

\textsc{CacheForge} resolves these issues by embedding the LLM inside a tightly governed, hardware-aware generate-evaluate-refine loop. Syntax- and structure-aware prompting ensures that every candidate conforms to the ChampSim interface, while static feasibility checks reject hallucinated or unsafe code before simulation. Storage-budget constraints and metadata accounting prevent unbounded growth, forcing the model to trade capacity against performance. A trace-based evaluator provides the causal feedback that imitation-based methods lack, punishing designs that increase hit rate but lengthen exposed stalls. Reward shaping aligns optimization with IPC and IPC-per-KB rather than raw hit rate, steering generation toward demand-critical behaviors. A three-tier memory system: ephemeral, episodic, and summary, retains useful design lessons without collapsing exploration, while dynamic temperature and mutation scheduling maintain diversity while encouraging novelty and prevent local minima. Structural crossover between top-performing but dissimilar policies injects genuine novelty, and model switching breaks late-stage stagnation. Together, these mechanisms convert an otherwise unstable generator into a disciplined architectural search engine capable of producing hardware friendly, feasible, and high-performance replacement algorithms.


A further limitation of prior learning-based replacement policies is the complete absence of \textit{explainability} and the lack of phase- or workload-awareness. PARROT\cite{parrot} and Glider\cite{glider} attempt to interpret their neural predictors by inspecting attention vectors or LSTM activations, and CHiRP\cite{chirp} analyzes perceptron and Adaline weights to identify correlated features; however, these signals reflect only statistical associations learned \emph{offline} from fixed training traces. Once training ends, the model’s behavior is frozen: it cannot adapt its interpretation of locality across phases, workloads, or architectural contexts, nor can it reveal the causal reasoning behind its eviction choices. At best, these frameworks expose coarse indicators of correlation, e.g., which PCs were influential in the training distribution, but not the algorithmic principles governing online decision-making. This has significantly limited the use of advanced AI in the field and the \textit{trust} level needed for implementing these policies in the hardware. 

In contrast, \textsc{CacheForge} represents each policy as executable C++ code, making its mechanisms, metadata structures, features, and decision making, explicitly visible and analyzable. Because evolution occurs inside a trace-based simulator, every modification is validated against the \emph{current} phase behavior and workload dynamics, not a frozen offline trace. Moreover, each evolutionary step is archived in a persistent memory ledger, allowing users to inspect how the policy’s logic changes over time, which structures emerge or vanish, and which design ideas survive selection. This yields the first explainable learning-driven replacement framework: the causal logic of the policy is transparent because it is written in code, and its developmental trajectory is transparent because every mutation, crossover, and rationale is recorded. Prior neural approaches reveal correlations after training; \textsc{CacheForge} reveals the algorithm as it is created.

The remainder of this paper formalizes this approach and demonstrates that governed code-space evolution can synthesize compact, high-performance replacement mechanisms beyond the reach of fixed representations.

\noindent\textbf{Contributions}. This paper makes four contributions:
\begin{itemize}
 \item \textbf{Generative code-space evolution for cache replacement.}  
We introduce \textsc{CacheForge}, the first system that treats a replacement policy as \emph{mutable code}, allowing the LLM to rewrite control flow, metadata structures, decision rules, and scoring logic, while a trace-based ChampSim evaluator provides causal feedback and enforces feasibility. This moves beyond fixed representations such as IPV vectors, RRIP templates, or MPPPB feature libraries, enabling algorithmic forms no prior framework can express autonomously. 

\item \textbf{Autonomous discovery of high-performance and hardware-efficient mechanisms.}  
Governed evolution synthesizes compact policies that surpass all CRC-2 baselines. Across SPEC CPU2006, \textsc{CacheForge} improves IPC by {10.15\%}, {7.89\%}, {6.34\%}, {3.64\%}, {3.12\%}, and {2.71\%} over LRU, MPPPB, LiMe, ReD, SHiP++, and Hawkeye, and raises total hit rate by {27.36\%}, {19.69\%}, {13.72\%}, {13.15\%}, {11.83\%}, and {5.73\%} over MPPPB, ReD, Hawkeye, SHiP++, LiMe, and LRU. It simultaneously reduces metadata overhead by {43\%} relative to MPPPB and delivers the highest IPC-per-KB efficiency among all evaluated designs. 


\item \textbf{Comprehensive analysis of novel policies and sensitivity study.}  
We identify 21 novel design components and several algorithmic hybrids synthesized only through code-space evolution, including 
SPARC+, TA-DPIP, RSP-MPPP, and SAGE-MPPPB. Sensitivity studies show how memory, temperature, mutation rate, prompt, and seed diversity shape convergence among 10,687 generated policies.  
\end{itemize}

\noindent \textbf{Paper Organization}: Section~\ref{sec:background} provides background and related work. 
Section~\ref{sec:design} presents the \textsc{CacheForge} design. 
Section~\ref{sec:methodology} describes our evaluation methodology. 
Section~\ref{sec:results} reports experimental results. 
Section~\ref{sec:sensitivity} provides sensitivity analysis. 
Section~\ref{sec:conclusion} concludes the paper.

\section{Background and Related Works}
\label{sec:background}

\noindent\paragraph{Classical heuristic policies}
Classical LLC replacement mechanisms such as DIP~\cite{dip}, RRIP~\cite{rrip}, and SHiP~\cite{ship} are manually engineered around recency, re-reference interval prediction, or lightweight program-context signatures. These heuristics are robust, interpretable, and low cost, but the decision logic is fixed: their metadata structures, state transitions, and victim-selection rules are predetermined and cannot evolve. ReD\cite{red} inserts a line only after observing a prior reuse and uses PC-based logic to bypass predicted-dead blocks. SHiP++\cite{ship++} refines signature-based reuse prediction with improved SHCT training and separate handling of demand and prefetch lines. MPPPB\cite{mpppb} uses a perceptron that combines multiple program features to estimate reuse likelihood for bypass and insertion decisions. 

\noindent\paragraph{Imitation-based learning}
Recent approaches attempt to approximate Belady or exploit rich locality structure using offline learning. LIME\cite{less-is-more} trains a lightweight demand-access classifier to approximate Belady and uses the prediction to choose insertion depth under SRRIP. Hawkeye~\cite{hawkeye} constructs per-PC labels using Belady's MIN algorithm and learns a lightweight classifier that distinguishes ``cache-friendly'' from ``cache-averse'' lines. Glider~\cite{glider} trains deep sequence models to capture long-range reuse patterns and distills them into a compact online predictor. PARROT~\cite{parrot} uses LSTM-based sequence modeling to imitate Belady's optimal algorithm~\cite{belady} directly from traces. These methods improve pattern recognition but share two fundamental limitations:  
(1) their algorithmic structure is frozen after training—no new metadata, control flow, or decision logic can emerge; and  
(2) they operate only on correlation signals, not the true causal impact of a decision on future execution.

\noindent\paragraph{Evolutionary frameworks with fixed representations}
IPV~\cite{ipv} encodes each policy as an insertion–promotion vector and uses a genetic algorithm to search this space. Although this includes causal evaluation via simulation, the representation restricts behavior to LRU-stack manipulations; IPV cannot invent new mechanisms or metadata behaviors.  
MPPPB~\cite{mpppb} aggregates information from sixteen expert-designed perspectives (PC bits, region signatures, reuse classifications, temporal decay signals, access-type discriminators, and more), exploring over $10^{110}$ possible feature combinations. This search required roughly ten CPU-years and produced strong predictors, but the algorithmic form, feature extraction followed by weighted summation and thresholding is fixed.  
PACIPV~\cite{pacipv} restricts exploration to $1024$ RRIP-derived vectors, enabling filtered exhaustive search, but the search is confined to the RRIP abstraction and yields only marginal IPC gains.  
These frameworks employ causal evaluation but are limited to the behaviors representable within their fixed templates.

\noindent\paragraph{Foundational LLM Reasoning Frameworks}
LLMs have shown promising results when it comes to solving multi-step problems, by leveraging techniques such as Chain-of-Thought prompting~\cite{cot} and zero-shot reasoning~\cite{zero-shot-reasoning}. Their reasoning capabilities were further augmented with agent-based frameworks like ReAct~\cite{react}, which combines reasoning with environment-specific actions.
Building on these LLM capabilities, a new class of frameworks has emerged that focus on automated optimization and discovery. These frameworks deal with using LLMs to optimize their own prompts, such as APE~\cite{ape} and PromptWizard~\cite{prompt-wizard}, automatically generate autonomous agents with frameworks like Autoagents~\cite{autoagents}, and discovering novel algorithms for producing ``as-good-as-possible'' solutions to NP-Hard problems (FunSearch)~\cite{funsearch_nphard} and even finding solutions to various other scientific and mathematical problems (AlphaEvolve)~\cite{alphaevolve}. Recent studies have shown that LLMs are prone to ``hallucinations'' resulting in production of confident falsehoods~\cite{hallucinations} and lack causal, domain-specific grounding, often limiting their ability to be applicable to the hard constraints of physical hardware design.

\section{\textsc{CacheForge}}
\label{sec:design}



\begin{figure*}
    \centering
    \includegraphics[width=\textwidth]{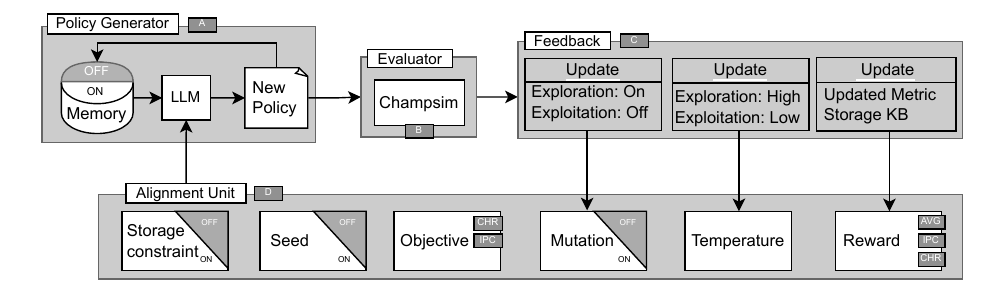}
    \caption{\textsc{CacheForge} system diagram.}
    \label{fig:pipeline}
\end{figure*}




Imitation-based frameworks capture statistical correlations but cannot adapt causally or revise their structure. Causal approaches incorporate simulation feedback but remain limited by fixed, human-designed representations. A more flexible framework is needed, one that supports code generation, causal feedback, hardware-aware objectives, and generalization across workloads.

Prior evaluators also suffer from misaligned objectives. Most optimize hit rate or MPKI, assuming a monotonic relationship with IPC. This assumption breaks down in modern processors: demand and RFO misses stall retirement, whereas prefetch hits, writebacks, and speculative accesses raise hit rate without improving performance. As a result, policies such as SHiP, RRIP, MPPPB, and PACIPV often increase total hits while reducing demand-service efficiency. A mechanism to tune the reward and optimization target using direct feedback is therefore necessary.

\textsc{CacheForge} addresses this gap by enabling evolutionary exploration of executable replacement-policy code under true hardware constraints and causal feedback.

\textsc{CacheForge} embeds an LLM inside a closed generator-evaluator loop. At each iteration, the generator proposes candidate code, ChampSim evaluates it under identical setups, and a reward guides refinement. The evaluator enforces physical correctness, preventing hallucinated or semantically incorrect code from propagating. 


\noindent\textbf{{Policy Generator.}}
The Policy Generator (Figure~\ref{fig:pipeline} A) connects an LLM to the CRC-2 ChampSim loop to autonomously synthesize C++ replacement policies. Each iteration retrieves top-performing policies, their metadata, and rationales from a policy database. 
These are encoded with objectives, reward, constraints, and workload summaries into a structured prompt. The LLM produces a policy name, short rationale, C++ implementation, and explicit overhead estimate.

A strict one-shot prompting schema ensures syntactic validity and compatibility with ChampSim, while high-performing policies are reintroduced in subsequent prompts to enable chain-of-thought–style refinement. 
\textsc{CacheForge} also supports dynamic seeding: generation may begin from example baselines such as RRIP, SHiP, Hawkeye, or MPPPB for guided refinement, or from an empty context for free-form exploration. This flexibility allows the system to alternate naturally between anchored optimization and open-ended innovation.

\noindent\textbf{Simulation \& Evaluation.}
Each generated policy is compiled using the CRC-2 version of ChampSim and executed across representative SPEC CPU2006 workloads. The simulator returns performance metrics such as IPC, per-access-type hit rates and MPKI, and estimated storage overhead, which are automatically parsed and recorded (Figure~\ref{fig:pipeline} B). These quantitative results serve as the feedback signal that steers the next design iteration, transforming the system into a self-contained reinforcement loop.

\noindent\textbf{Objective and Reward.}
Reward translates simulator statistics into a single scalar signal that drives the \textsc{CacheForge} optimization loop. After each generated policy is simulated, ChampSim reports key cache metrics like \textit{IPC}, \textit{MPKI}, \textit{total hit rate}, and per-access-type hit rates (\textit{Load}, \textit{RFO}, \textit{Prefetch}, and \textit{Writeback}). These metrics are combined according to the active optimization objective defined by the Optimization Objective Controller (Figure~\ref{fig:pipeline} C,D).

\textsc{CacheForge} distinguishes between the \emph{reward} returned by simulation and the \emph{objective} embedded in the prompt. This separation enables guided divergence, encouraging innovation rather than premature convergence.

Moreover, the reward serves both as evaluation and as alignment: it penalizes excessive metadata, latency-increasing structures, or useless prefetch retention by instructing the LLM to track storage overhead, constraining designs to remain physically realizable and comparable to existing CRC-2 implementations in both capacity and complexity. Finally, high-reward policies populate the positive memory set; low-reward ones enter negative conditioning.

\noindent\textbf{Memory-Guided Refinement.} \textsc{CacheForge} maintains a three-tier memory hierarchy to accumulate and reuse design knowledge. \textit{Ephemeral memory} captures short-term context from the current round. \textit{Episodic memory} stores structured experiment outcomes and short textual lessons. \textit{Summary memory} distills persistent principles, constraints, and “don’ts” to inform future prompts.
Before each new iteration, the system retrieves the most relevant memories using vector similarity and prepends them to the prompt, ensuring continuity of reasoning. This mechanism allows \textsc{CacheForge} to avoid redundant exploration and progressively internalize its own design insights  (Figure~\ref{fig:pipeline} A).

\noindent\textbf{Exploration–Exploitation.}
\textsc{CacheForge} operates in two phases: \emph{exploration}, where the system searches broadly for new algorithmic structures, and \emph{exploitation}, where it refines high-performing designs. The system transitions between these phases based on recent simulation results: lack of improvement triggers exploration, while consistent gains shift the system toward exploitation  (Figure~\ref{fig:pipeline} C).

To implement these phases, \textsc{CacheForge} uses three lightweight control mechanisms.  
First, a \textit{Dynamic Temperature Controller (DTC)} adjusts the model’s sampling temperature~($\tau$). High temperatures ($\tau = 0.9$--1.0) promote diversity during exploration, whereas lower temperatures ($\tau = 0.7$--0.8) stabilize refinement during exploitation. Empirically, moderate values near $\tau = 0.7$ strike an effective balance between novelty and convergence.

Second, a \textit{mutation controller} determines how strongly each new policy should deviate from its predecessor. Improvements trigger small refinements (mutation rate 0.2), moderate regressions increase mutation to 0.4, and large regressions raise it to 0.9 or 1.0 to force substantial structural changes. Plateaus induce a moderate mutation rate (0.6) to escape stagnation. These directives guide the degree of permissible algorithmic change each iteration.

Finally, when refinement no longer yields progress, \textsc{CacheForge} can activate a \textit{tree-based diversification}. The system (i) further refines the current parent policy, (ii) performs crossover between top-performing but structurally different designs, and (iii) introduces entirely new policies using accumulated design lessons. This branching search helps escape local optima and reintroduces structural diversity when linear refinement is insufficient.

Together, temperature adaptation, mutation control, and tree-based diversification maintain broad exploration early in search and support focused, stable convergence once promising design directions emerge.


\section{Methodology}
\label{sec:methodology}



\noindent\textbf{Simulator and Workload.}  We use the CRC2-ChampSim trace-based simulator \cite{champsim} to evaluate policies for single core configuration with prefetcher both enabled \& disabled. Each policy is integrated into the CRC2-ChampSim codebase and evaluated using 28 SPEC 2006 CPU traces. We report hit rate, IPC, Speedup, and IPC-per KB for all workloads and as well as the memory-intensive subset of five workloads (astar, mcf, milc, lbm, omnetpp) with the highest sorted miss rate.  We report the hardware overhead and our best IPC-per-KB policy for all 88 simulation points after 200M warmup and 1B instructions, 
execution. 
\begin{table}[!t]
\footnotesize
\centering
\caption{Processor and Memory Configuration.
}
\label{tab:sim_params}
\begin{tabularx}{\columnwidth}{@{}lX@{}}
\toprule
\textbf{Component} & \textbf{Configuration} \\
\midrule
Processor & 1 core; 4 GHz; 6-wide fetch/decode/execute; 4-wide retire; 256-entry ROB; gshare branch predictor \\
L1 I/D-Cache & 32 KB, 64 sets, 8 ways, 4-cycle latency, next-line prefetcher;  \\ 
L2 Cache & 256 KB, 512 sets, 8 ways; 12-cycle latency, PC-based stride prefetcher; \\ 
LLC & 2 MB, 2048 sets, 16 ways; 32-cycle latency;  \\
L1 TLB (I/D) & 4KB; 16 sets, 8 ways \\
L2 TLB & 96KB; 128 sets, 12 ways \\
\bottomrule
\end{tabularx}
\end{table}

\textsc{CacheForge} automates the entire code–evaluation pipeline with zero-human intervention through the evolutionary phase. 
Each iteration consists of four stages:  
(1) \textit{Generation}, where the LLM produces a syntactically valid C++ policy;  
(2) \textit{Compilation}, which integrates the policy into ChampSim;  
(3) \textit{Execution}, where the policy is simulated under each evaluation window; and  
(4) \textit{Evaluation}, where key metrics like IPC, total hit rate, per-access-type hit rates, MPKI, and estimated metadata cost are extracted.  
This study compares the results of  10,687 fully autonomously generated policies.

We compare results for the best policy over 260 iterations of \textsc{CacheForge} evolution. We observed that evolving \textsc{CacheForge} by evaluating the workloads for 100M vs 1B does not change the \textit{ranking} of the policies but significantly reduces the evaluation overhead for \textsc{CacheForge} per iteration required for evolution.  We provide insights on \textsc{CacheForge's} evolution for 10\,M(1 M warmup), 100\,M(100 M warmup), and 1\,B(200 M warmup) execution per instruction since it affects the total evolutionary run time and explain how it impacts the final hit rate, IPC, IPC improvement Per-KB. 

\noindent\textbf{Baseline Replacement Policies.} 
We compare the best \textsc{CacheForge} generated policies with the top-performing CRC2 replacement policies: SHiP++, Hawkeye, Less Is More (LIME), Multiperspective (MPPPB), Reuse Detection (ReD), and LRU, as  \textit{human-designed} seed examples \cite{ship, hawkeye,mpppb}. 
These perform the impact of each of these policies provided to the \textsc{CacheForge} as seeded runs and report its impact on the evolution of CacheForge quantitatively. 
We consider MPPPB the state of the art AI-assisted  replacement policy generator and compare \textsc{CacheForge} novel feature generation with that. 

\noindent\textbf{Language Models Evaluated.}
We evaluate \textsc{CacheForge} using multiple large language models under identical prompt schemas and hyperparameters:  
\texttt{GPT-5}, \texttt{o4-mini}, \texttt{Gemini 2.5 Pro}, \texttt{DeepSeek-Coder 33B}, and \texttt{Qwen 2.5}.  
This comparison isolates architectural effects from model-specific generative behavior.

\vspace{-0.25em}

\section{Results}

\label{sec:results}

\subsection{Guided Evolution for Various Objectives.}
This section shows how \textsc{CacheForge} can be used to optimize various objectives. Including metrics with complex dynamic interactions in CPU, such as hit rate and IPC. 

Although LLC hit rate is often treated as a proxy for performance, in practice it is a remarkably poor stand-in for what the core actually experiences on its critical path, and a naive attempt to “maximize hit rate” can easily degrade IPC. 

The fundamental problem is that hit rate is an undifferentiated aggregate over access types, sets, and phases, whereas stalls are driven by a tiny subset of accesses: long-latency, poorly overlapped demand loads. A replacement policy can raise total hit rate by zealously protecting prefetch and writeback lines, or by retaining demand lines in cold sets, yet simultaneously evict the very lines that matter in hot sets where reuse is dense and MLP is limited. 

From a statistical point of view, a prefetch that hits looks identical to a load that hits; from the processor’s point of view, the former is often invisible while the latter may save tens of cycles in a dependence chain. Even worse, the rebalancing of occupancy that improves the global hit rate can reduce memory-level parallelism or increase bank and MSHR contention, making the remaining misses more serialized and more frequently exposed to the retire stage. The result is that we can construct policies with strictly higher hit rate and lower MPKI that nonetheless lengthen critical-path stalls and reduce IPC. In other words, maximizing hit rate optimizes for “how many lines live in the cache”, not for “which misses the core actually pays for”. 


\subsection{Automatically Learning What to Optimize}

\begin{figure}[!htbp]
\centering
\begin{tikzpicture}

\begin{axis}[
    width=7.5cm,
    height=4.5cm,
    xlabel={Sorted Policy},
    ylabel={IPC},
    xmin=1, xmax=90,
    ymin=0.30, ymax=0.34,
    ymajorgrids=true,
    grid style={dotted,gray!40},
    axis y line*=left,
    axis x line*=bottom,
    tick label style={font=\scriptsize},
    label style={font=\small},
]

\addplot[
    color=blue!80!black,
    very thick,
    mark=none
] coordinates {
(1,0.333264)(2,0.332885)(3,0.332482)(4,0.332032)(5,0.331937)(6,0.331880)
(7,0.331708)(8,0.331617)(9,0.331597)(10,0.331385)(11,0.331268)(12,0.331238)
(13,0.331230)(14,0.331197)(15,0.330960)(16,0.330879)(17,0.330756)(18,0.330749)
(19,0.330739)(20,0.330710)(21,0.330696)(22,0.330665)(23,0.330580)(24,0.330312)
(25,0.330200)(26,0.329815)(27,0.329404)(28,0.329385)(29,0.327954)(30,0.327847)
(31,0.327810)(32,0.327804)(33,0.327802)(34,0.327566)(35,0.327556)(36,0.326322)
(37,0.324392)(38,0.324288)(39,0.323935)(40,0.323768)(41,0.323279)(42,0.322865)
(43,0.322718)(44,0.322522)(45,0.322086)(46,0.321983)(47,0.321058)(48,0.320961)
(49,0.320527)(50,0.318561)(51,0.318520)(52,0.316672)(53,0.316610)(54,0.316609)
(55,0.316310)(56,0.315859)(57,0.315719)(58,0.315440)(59,0.314875)(60,0.314263)
(61,0.314216)(62,0.313934)(63,0.313852)(64,0.313774)(65,0.313561)(66,0.313551)
(67,0.313191)(68,0.313170)(69,0.312944)(70,0.312754)(71,0.312298)(72,0.312169)
(73,0.312150)(74,0.311633)(75,0.311489)(76,0.310916)(77,0.310781)(78,0.310727)
(79,0.309508)(80,0.307987)(81,0.307291)(82,0.305721)(83,0.304521)(84,0.303268)
(85,0.302368)(86,0.301704)(87,0.301585)(88,0.301431)(89,0.301375)(90,0.301234)
};

\end{axis}

\begin{axis}[
    width=7.5cm,
    height=4.5cm,
    xmin=1, xmax=90,
    ymin=0.30, ymax=0.50,
    ylabel={Cache Hit Rate},
    axis y line*=right,
    axis x line=none,
    tick label style={font=\scriptsize},
    label style={font=\small},
    yticklabel style={text width=2em,align=right},
]

\addplot[
    forget plot,
    only marks,
    mark=*,
    mark size=1pt,
    color=blue!60!cyan,
    opacity=0.5
]
coordinates {
(1,0.3698)(2,0.3986)(3,0.3653)(4,0.3867)(5,0.3673)(6,0.3666)(7,0.3828)(8,0.3697)
(9,0.4000)(10,0.3673)(11,0.3798)(12,0.3858)(13,0.3577)(14,0.3597)(15,0.3554)
(16,0.3553)(17,0.3567)(18,0.4029)(19,0.3565)(20,0.3976)(21,0.3564)(22,0.3568)
(23,0.4082)(24,0.3722)(25,0.3695)(26,0.3813)(27,0.4025)(28,0.3770)(29,0.3737)
(30,0.3915)(31,0.3759)(32,0.4059)(33,0.3649)(34,0.3877)(35,0.4001)(36,0.4010)
(37,0.3934)(38,0.4100)(39,0.3840)(40,0.4145)(41,0.3937)(42,0.4282)(43,0.3983)
(44,0.4156)(45,0.3626)(46,0.4251)(47,0.3984)(48,0.3594)(49,0.4651)(50,0.4154)
(51,0.4202)(52,0.4659)(53,0.4605)(54,0.4604)(55,0.4469)(56,0.4555)(57,0.4569)
(58,0.4680)(59,0.4765)(60,0.4835)(61,0.4713)(62,0.4840)(63,0.4823)(64,0.4416)
(65,0.4409)(66,0.4763)(67,0.3207)(68,0.4532)(69,0.4569)(70,0.4277)(71,0.4951)
(72,0.4771)(73,0.4954)(74,0.4829)(75,0.4882)(76,0.4758)(77,0.4804)(78,0.4887)
(79,0.4912)(80,0.4227)(81,0.4760)(82,0.4610)(83,0.4331)(84,0.4367)(85,0.4302)
(86,0.4273)(87,0.4256)(88,0.4147)(89,0.4153)(90,0.4207)
};


\end{axis}

\end{tikzpicture}
\caption{Anti-correlation of IPC and Cache Hit Rate observed.}
\label{fig:ipc_hit_dual_axis}
\end{figure}
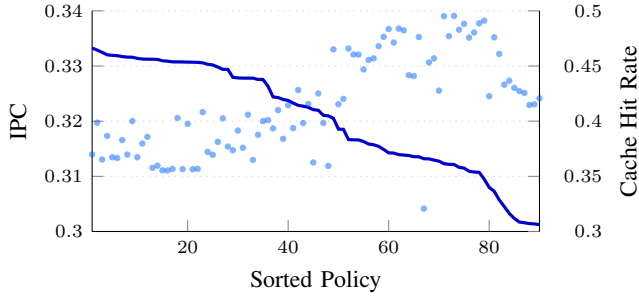

Our first and most surprising observation is that, once we let \textsc{CacheForge} run long enough, it immediately disproves the naive objective we started with. If we sort the thousands of generated policies by LLC hit rate and then examine their IPC, we often see the \emph{opposite} of what a human designer would expect: many of the highest-hit-rate policies deliver \emph{lower} IPC than the baselines, while some of the best-IPC policies live in the middle of the hit-rate distribution (Figure~\ref{fig:ipc_hit_dual_axis}). 

From a traditional, hand-designed study with a handful of policies, this is easy to miss or dismiss as an anomaly. In \textsc{CacheForge}, the pattern is impossible to ignore, because the generator–evaluator loop continuously produces entire populations of policies that cover a wide swath of the design space. The system effectively runs the thought experiment a human architect would find exhausting: “what happens to IPC if I push hit rate up in every conceivable way?”; and the answer, repeatedly, is that aggregate hit rate is the wrong hill to climb.



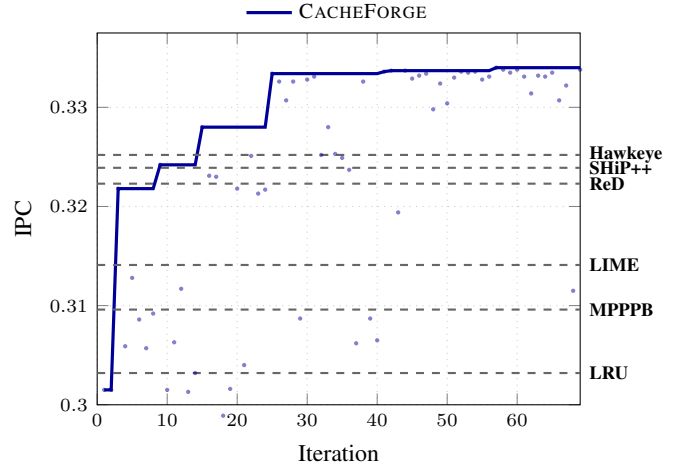
\begin{figure}[!h]
\centering
\begin{tikzpicture}
\begin{axis}[
    width=0.9\columnwidth,
    height=6.5cm,
    xlabel={Iteration},
    ylabel={IPC},
    xmin=0, xmax=69,
    ymin=0.30, ymax=0.3375,
    grid=both,
    grid style={dotted,gray!40},
    legend style={
        font=\footnotesize,
        at={(0.5,1)}, 
        anchor=south,
        draw=none,
        fill=none,
    },
    legend columns=-1,
    tick label style={font=\scriptsize},
    label style={font=\small},
    legend cell align=left,
    clip=false
]

\addplot[blue!60!black, very thick, mark=none] coordinates {
(1,0.3015)(2,0.3015)(3,0.3218)(4,0.3218)(5,0.3218)(6,0.3218)(7,0.3218)(8,0.3218)
(9,0.3242)(10,0.3242)(11,0.3242)(12,0.3242)(13,0.3242)(14,0.3242)
(15,0.3280)(16,0.3280)(17,0.3280)(18,0.3280)(19,0.3280)(20,0.3280)
(21,0.3280)(22,0.3280)(23,0.3280)(24,0.3280)
(25,0.3334)(26,0.3334)(27,0.3334)(28,0.3334)(29,0.3334)(30,0.3334)
(31,0.3334)(32,0.3334)(33,0.3334)(34,0.3334)(35,0.3334)(36,0.3334)(37,0.3334)
(38,0.3334)(39,0.3334)(40,0.3334)
(41,0.3336)(42,0.3337)(43,0.3337)(44,0.3337)(45,0.3337)(46,0.3337)
(47,0.3337)(48,0.3337)(49,0.3337)(50,0.3337)
(51,0.3337)(52,0.3337)(53,0.3337)(54,0.3337)(55,0.3337)(56,0.3337)
(57,0.3340)(58,0.3340)(59,0.3340)(60,0.3340)
(61,0.3340)(62,0.3340)(63,0.3340)(64,0.3340)(65,0.3340)(69,0.3340)
};\addlegendentry{\textsc{CacheForge}}

\addplot[only marks, mark=*, mark size=0.6pt, color=blue!60!black, opacity=0.5]
coordinates {
(1,0.3015)(2,0.3015)(3,0.3218)(4,0.3059)(5,0.3128)(6,0.3086)(7,0.3057)
(8,0.3092)(9,0.3242)(10,0.3015)(11,0.3063)(12,0.3117)(13,0.3013)(14,0.3032)
(15,0.3280)(16,0.3231)(17,0.3230)(18,0.2989)(19,0.3016)(20,0.3218)
(21,0.3040)(22,0.3251)(23,0.3213)(24,0.3217)(25,0.3334)(26,0.3326)
(27,0.3307)(28,0.3326)(29,0.3087)(30,0.3328)(31,0.3331)(32,0.3252)
(33,0.3280)(34,0.3253)(35,0.3249)(36,0.3237)(37,0.3062)(38,0.3326)
(39,0.3087)(40,0.3065)(41,0.3336)(42,0.3337)(43,0.3194)(44,0.3337)
(45,0.3329)(46,0.3332)(47,0.3334)(48,0.3298)(49,0.3324)(50,0.3304)
(51,0.3330)(52,0.3336)(53,0.3335)(54,0.3336)(55,0.3328)(56,0.3331)
(57,0.3340)(58,0.3338)(59,0.3335)(60,0.3338)(61,0.3331)(62,0.3314)
(63,0.3332)(64,0.3331)(65,0.3335)(66,0.3307)(67,0.3322)(68,0.3115)(69,0.3338)
};

\addplot[dashed, black!60, thick, domain=0:69]{0.3032};
\addplot[dashed, black!60, thick, domain=0:69]{0.3096};
\addplot[dashed, black!60, thick, domain=0:69]{0.3141};
\addplot[dashed, black!60, thick, domain=0:69]{0.3223};
\addplot[dashed, black!60, thick, domain=0:69]{0.3239};
\addplot[dashed, black!60, thick, domain=0:69]{0.3252};

\node[anchor=west,font=\scriptsize\bfseries,text=black]
    at (axis cs:69,0.3032){LRU};
\node[anchor=west,font=\scriptsize\bfseries,text=black]
    at (axis cs:69,0.3096){MPPPB};
\node[anchor=west,font=\scriptsize\bfseries,text=black]
    at (axis cs:69,0.3141){LIME};
\node[anchor=west,font=\scriptsize\bfseries,text=black]
    at (axis cs:69,0.3223){ReD};
\node[anchor=west,font=\scriptsize\bfseries,text=black]
    at (axis cs:69,0.3239){SHiP++};
\node[anchor=west,font=\scriptsize\bfseries,text=black]
    at (axis cs:69,0.3252){Hawkeye};


\end{axis}
\end{tikzpicture}
\caption{Optimizing for IPC over iterations. } 

\label{fig:ipc_evolution}
\end{figure}

\begin{figure}[!htbp]
\centering
\begin{tikzpicture}
\begin{axis}[
    width=\columnwidth,
    height=3.5cm,
    ybar,
    bar width=3pt,
    ylabel={Speedup(\%)},
    xlabel={Workload},
    symbolic x coords={mcf,lbm,astar,omnetpp,milc,Average},
    xtick={lbm,omnetpp,milc,astar,mcf,Average},
    xticklabel style={rotate=45,
    anchor=east,font=\footnotesize},
    ymajorgrids=true,
    grid style={dotted,gray!40},
    enlarge x limits=0.12,
    ymin=0,
    ymax=20,
    tick label style={font=\footnotesize},
    label style={font=\normalsize},
    legend style={
        font=\footnotesize,
        at={(0.5,1.05)},
        anchor=south,
        legend columns=-1,
        draw=none
    },
    legend entries={CacheForge,Hawkeye,SHiP++,ReD,LiMe},
]

\addplot+[ybar, bar shift=-8pt, fill=blue!90!black!60, draw=black] coordinates {(lbm,11.4220)}; 
\addplot+[ybar, bar shift=-4pt,  fill=white, draw=black] coordinates {(lbm,10.7551)};           
\addplot+[ybar, bar shift=0pt, fill=gray!45, draw=black!80, pattern=north east lines] coordinates {(lbm,9.6802)};       
\addplot+[ybar, bar shift=+4pt, fill=gray!85, draw=black!80, pattern=crosshatch] coordinates {(lbm,11.3361)};      
\addplot+[ybar, bar shift=+8pt, fill=gray!65, draw=black!70, pattern=horizontal lines] coordinates {(lbm,3.0849)};       

\addplot+[ybar, bar shift=-8pt, fill=blue!90!black!60, draw=black] coordinates {(omnetpp,6.4766)}; 
\addplot+[ybar, bar shift=-4pt,  fill=white, draw=black] coordinates {(omnetpp,4.3213)};             
\addplot+[ybar, bar shift=0pt, fill=gray!45, draw=black!80, pattern=north east lines] coordinates {(omnetpp,5.2044)};        
\addplot+[ybar, bar shift=+4pt, fill=gray!85, draw=black!80, pattern=crosshatch] coordinates {(omnetpp,0.0292)};        
\addplot+[ybar, bar shift=+8pt, fill=gray!65, draw=black!70, pattern=horizontal lines] coordinates {(omnetpp,3.4962)};        

\addplot+[ybar, bar shift=-8pt, fill=blue!90!black!60, draw=black] coordinates {(milc,11.2966)}; 
\addplot+[ybar, bar shift=-4pt, fill=white, draw=black] coordinates {(milc,3.8694)};             
\addplot+[ybar, bar shift=0pt, fill=gray!45, draw=black!80, pattern=north east lines] coordinates {(milc,2.4299)};        
\addplot+[ybar, bar shift=+4pt,  fill=gray!85, draw=black!80, pattern=crosshatch] coordinates {(milc,3.0308)};        
\addplot+[ybar, bar shift=+8pt, fill=gray!65, draw=black!70, pattern=horizontal lines] coordinates {(milc,1.1098)};        

\addplot+[ybar, bar shift=-8pt, fill=blue!90!black!60, draw=black] coordinates {(astar,11.1119)}; 
\addplot+[ybar, bar shift=-4pt,  fill=white, draw=black] coordinates {(astar,7.9153)};             
\addplot+[ybar, bar shift=0pt, fill=gray!45, draw=black!80, pattern=north east lines] coordinates {(astar,6.9883)};        
\addplot+[ybar, bar shift=+4pt, fill=gray!85, draw=black!80, pattern=crosshatch] coordinates {(astar,10.6608)};        
\addplot+[ybar, bar shift=+8pt, fill=gray!65, draw=black!70, pattern=horizontal lines] coordinates {(astar,6.4921)};        

\addplot+[ybar, bar shift=-8pt, fill=blue!90!black!60, draw=black] coordinates {(mcf,16.7288)}; 
\addplot+[ybar, bar shift=-4pt, fill=white, draw=black] coordinates {(mcf,15.7602)};            
\addplot+[ybar, bar shift=0pt, fill=gray!45, draw=black!80, pattern=north east lines] coordinates {(mcf,17.7933)};       
\addplot+[ybar, bar shift=+4pt,  fill=gray!85, draw=black!80, pattern=crosshatch] coordinates {(mcf,17.7041)};      
\addplot+[ybar, bar shift=+8pt, fill=gray!65, draw=black!70, pattern=horizontal lines] coordinates {(mcf,16.8853)};      

\addplot+[ybar, bar shift=-8pt, fill=blue!90!black!60, draw=black] coordinates {(Average,10.1861)}; 
\addplot+[ybar, bar shift=-4pt, fill=white, draw=black] coordinates {(Average,7.2764)};             
\addplot+[ybar, bar shift=0pt,  fill=gray!45, draw=black!80, pattern=north east lines] coordinates {(Average,6.8343)};        
\addplot+[ybar, bar shift=+4pt, fill=gray!85, draw=black!80, pattern=crosshatch] coordinates {(Average,6.3185)};        
\addplot+[ybar, bar shift=+8pt, fill=gray!65, draw=black!70, pattern=horizontal lines] coordinates {(Average,3.6021)};        

\end{axis}
\end{tikzpicture}

\caption{Speedup of \textsc{CacheForge} relative to LRU across SPEC workloads, compared against state-of-the-art baselines.}
\label{fig:speedup_over_lru}
\end{figure}
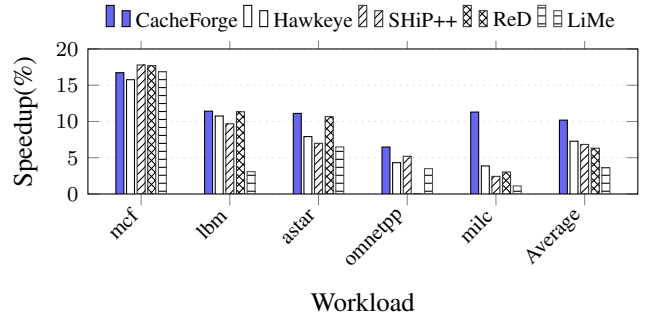

\noindent\textbf{IPC.}
Figure~\ref{fig:ipc_evolution} shows that IPC rises steadily over time for \textsc{CacheForge}. 
only 69 iterations of CacheForge improves IPC to
0.334, surpassing all SOTA baselines with improvements of 10.16\% over LRU, 7.89\% over
MPPPB, 6.34\% over LiMe, 3.64\% over ReD, 3.12\% over SHiP++, and
2.71\% over Hawkeye. 
Across the evaluated workloads, \textsc{CacheForge} achieves substantial performance gains over LRU, improving IPC by 11.42\% on \texttt{lbm}, 6.48\% on \texttt{omnetpp}, 11.30\% on \texttt{milc}, 11.11\% on \texttt{astar}, and 16.73\% on \texttt{mcf}, with a 10.16\% speedup overall (Figure~\ref{fig:speedup_over_lru}). 


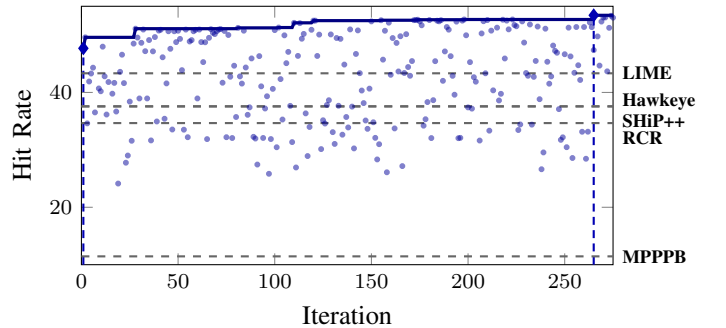
\begin{figure}[!htbp]
\centering
\begin{tikzpicture}
\begin{axis}[
    width=0.475\textwidth,
    height=5cm,
    xlabel={Iteration},
    ylabel={Hit Rate},
    xmin=0, xmax=275,          
    ymin=10, ymax=55,  
    y tick label style={
        /pgf/number format/fixed, 
        /pgf/number format/precision=3
    },
    tick label style={font=\footnotesize},
    label style={font=\normalsize},
    ylabel style={font=\normalsize},
    every axis plot/.append style={mark=none,line width=1pt},
    clip=false
]

\addplot[
only marks, mark=*, mark size=0.6pt, color=blue!60!black, opacity=0.5
]
coordinates {
(1,47.68)
(2,49.59)
(3,34.61)
(4,41.90)
(5,43.29)
(6,44.31)
(7,36.52)
(9,40.69)
(10,40.20)
(11,47.97)
(12,42.80)
(13,36.15)
(14,41.81)
(16,39.78)
(17,41.66)
(18,36.67)
(19,24.13)
(20,43.59)
(21,46.57)
(22,41.62)
(23,27.76)
(24,29.01)
(25,38.40)
(26,31.58)
(27,48.50)
(28,51.10)
(29,49.55)
(30,38.95)
(31,48.64)
(32,50.95)
(33,47.18)
(34,38.95)
(35,49.12)
(36,31.64)
(37,47.15)
(38,49.64)
(39,34.94)
(40,50.77)
(41,50.91)
(42,46.12)
(43,35.42)
(44,41.57)
(45,50.65)
(46,47.60)
(47,33.54)
(48,50.65)
(49,48.76)
(50,50.74)
(51,49.32)
(52,43.88)
(53,33.51)
(54,50.94)
(55,32.20)
(56,48.87)
(57,49.11)
(58,50.65)
(59,50.90)
(60,38.95)
(61,50.95)
(62,48.76)
(63,45.28)
(64,48.73)
(65,31.87)
(66,49.51)
(67,51.01)
(68,50.38)
(69,50.16)
(70,50.65)
(71,50.95)
(72,51.03)
(73,41.69)
(74,50.70)
(75,32.20)
(76,50.00)
(77,42.73)
(78,32.06)
(79,44.65)
(80,35.72)
(81,51.24)
(82,42.79)
(83,41.54)
(84,50.75)
(85,44.34)
(86,36.10)
(87,37.40)
(88,45.92)
(89,32.27)
(90,31.97)
(91,27.44)
(92,46.32)
(93,41.06)
(94,32.07)
(95,37.20)
(96,39.33)
(97,25.83)
(98,31.88)
(99,40.62)
(100,50.85)
(101,30.45)
(102,51.30)
(103,42.32)
(104,35.28)
(105,49.94)
(106,45.09)
(107,47.06)
(108,37.66)
(109,31.35)
(110,52.12)
(111,26.93)
(112,46.36)
(113,52.06)
(114,49.35)
(115,50.19)
(116,36.35)
(117,28.92)
(118,33.98)
(119,50.91)
(120,52.50)
(121,52.50)
(122,46.35)
(123,50.07)
(124,30.13)
(125,35.15)
(126,34.58)
(127,37.82)
(128,38.64)
(129,50.10)
(130,33.08)
(131,31.03)
(132,49.48)
(133,37.68)
(134,49.33)
(135,32.64)
(136,44.85)
(137,35.72)
(138,33.90)
(139,50.97)
(140,37.53)
(141,30.25)
(142,46.70)
(143,50.80)
(144,43.35)
(145,40.33)
(146,38.14)
(147,35.11)
(148,33.66)
(149,40.66)
(150,52.37)
(151,34.80)
(152,32.09)
(153,28.33)
(154,30.98)
(155,52.57)
(156,45.71)
(157,50.74)
(158,26.09)
(159,52.15)
(160,43.07)
(161,49.47)
(162,33.73)
(163,48.62)
(164,43.65)
(165,49.78)
(166,27.17)
(167,43.38)
(168,50.14)
(169,44.42)
(170,42.90)
(171,45.46)
(172,52.58)
(173,52.61)
(174,52.66)
(175,52.37)
(176,51.69)
(177,34.21)
(178,35.33)
(179,38.32)
(180,52.60)
(181,32.05)
(182,38.65)
(183,45.44)
(184,52.49)
(185,52.38)
(186,52.58)
(187,42.02)
(188,47.14)
(189,51.39)
(190,50.93)
(191,46.31)
(192,42.89)
(193,39.60)
(194,52.33)
(195,32.03)
(196,32.09)
(197,52.50)
(198,52.70)
(199,49.95)
(200,37.39)
(201,45.27)
(202,37.18)
(203,42.65)
(204,41.87)
(206,31.62)
(207,50.27)
(208,52.40)
(209,43.33)
(210,52.60)
(211,33.34)
(212,50.76)
(213,40.36)
(215,51.28)
(216,38.60)
(217,39.76)
(218,52.66)
(219,45.91)
(220,52.51)
(221,50.98)
(223,52.70)
(224,51.74)
(225,51.52)
(226,33.45)
(227,34.42)
(228,52.37)
(229,48.68)
(230,43.38)
(231,42.90)
(232,51.83)
(233,34.39)
(234,42.71)
(235,45.48)
(236,51.47)
(237,38.82)
(238,26.62)
(239,29.55)
(240,51.10)
(241,45.34)
(242,41.52)
(243,42.79)
(244,48.16)
(245,47.24)
(246,46.91)
(247,37.09)
(248,31.74)
(249,32.18)
(250,46.65)
(251,30.46)
(252,51.28)
(253,46.27)
(254,28.58)
(255,43.42)
(256,51.39)
(257,44.73)
(258,42.17)
(259,33.15)
(260,51.39)
(261,28.45)
(262,33.16)
(263,52.18)
(264,51.37)
(265,47.37)
(266,53.43)
(267,49.75)
(268,44.67)
(269,52.34)
(270,53.05)
(271,53.05)
(272,43.67)
(273,51.25)
(274,53.35)
(275,53.02)
};

\addplot[blue!50!black,very thick]
coordinates {
(1,47.68)
(2,49.59)
(3,49.59)
(4,49.59)
(5,49.59)
(6,49.59)
(7,49.59)
(8,49.59)
(9,49.59)
(10,49.59)
(11,49.59)
(12,49.59)
(13,49.59)
(14,49.59)
(15,49.59)
(16,49.59)
(17,49.59)
(18,49.59)
(19,49.59)
(20,49.59)
(21,49.59)
(22,49.59)
(23,49.59)
(24,49.59)
(25,49.59)
(26,49.59)
(27,49.59)
(28,51.10)
(29,51.10)
(30,51.10)
(31,51.10)
(32,51.10)
(33,51.10)
(34,51.10)
(35,51.10)
(36,51.10)
(37,51.10)
(38,51.10)
(39,51.10)
(40,51.10)
(41,51.10)
(42,51.10)
(43,51.10)
(44,51.10)
(45,51.10)
(46,51.10)
(47,51.10)
(48,51.10)
(49,51.10)
(50,51.10)
(51,51.10)
(52,51.10)
(53,51.10)
(54,51.10)
(55,51.10)
(56,51.10)
(57,51.10)
(58,51.10)
(59,51.10)
(60,51.10)
(61,51.10)
(62,51.10)
(63,51.10)
(64,51.10)
(65,51.10)
(66,51.10)
(67,51.10)
(68,51.10)
(69,51.10)
(70,51.10)
(71,51.10)
(72,51.10)
(73,51.10)
(74,51.10)
(75,51.10)
(76,51.10)
(77,51.10)
(78,51.10)
(79,51.10)
(80,51.10)
(81,51.24)
(82,51.24)
(83,51.24)
(84,51.24)
(85,51.24)
(86,51.24)
(87,51.24)
(88,51.24)
(89,51.24)
(90,51.24)
(91,51.24)
(92,51.24)
(93,51.24)
(94,51.24)
(95,51.24)
(96,51.24)
(97,51.24)
(98,51.24)
(99,51.24)
(100,51.24)
(101,51.24)
(102,51.30)
(103,51.30)
(104,51.30)
(105,51.30)
(106,51.30)
(107,51.30)
(108,51.30)
(109,51.30)
(110,52.12)
(111,52.12)
(112,52.12)
(113,52.12)
(114,52.12)
(115,52.12)
(116,52.12)
(117,52.12)
(118,52.12)
(119,52.12)
(120,52.50)
(121,52.50)
(122,52.50)
(123,52.50)
(124,52.50)
(125,52.50)
(126,52.50)
(127,52.50)
(128,52.50)
(129,52.50)
(130,52.50)
(131,52.50)
(132,52.50)
(133,52.50)
(134,52.50)
(135,52.50)
(136,52.50)
(137,52.50)
(138,52.50)
(139,52.50)
(140,52.50)
(141,52.50)
(142,52.50)
(143,52.50)
(144,52.50)
(145,52.50)
(146,52.50)
(147,52.50)
(148,52.50)
(149,52.50)
(150,52.50)
(151,52.50)
(152,52.50)
(153,52.50)
(154,52.50)
(155,52.57)
(156,52.57)
(157,52.57)
(158,52.57)
(159,52.57)
(160,52.57)
(161,52.57)
(162,52.57)
(163,52.57)
(164,52.57)
(165,52.57)
(166,52.57)
(167,52.57)
(168,52.57)
(169,52.57)
(170,52.57)
(171,52.57)
(172,52.58)
(173,52.61)
(174,52.66)
(175,52.66)
(176,52.66)
(177,52.66)
(178,52.66)
(179,52.66)
(180,52.66)
(181,52.66)
(182,52.66)
(183,52.66)
(184,52.66)
(185,52.66)
(186,52.66)
(187,52.66)
(188,52.66)
(189,52.66)
(190,52.66)
(191,52.66)
(192,52.66)
(193,52.66)
(194,52.66)
(195,52.66)
(196,52.66)
(197,52.66)
(198,52.70)
(199,52.70)
(200,52.70)
(201,52.70)
(202,52.70)
(203,52.70)
(204,52.70)
(205,52.70)
(206,52.70)
(207,52.70)
(208,52.70)
(209,52.70)
(210,52.70)
(211,52.70)
(212,52.70)
(213,52.70)
(214,52.70)
(215,52.70)
(216,52.70)
(217,52.70)
(218,52.70)
(219,52.70)
(220,52.70)
(221,52.70)
(222,52.70)
(223,52.70)
(224,52.70)
(225,52.70)
(226,52.70)
(227,52.70)
(228,52.70)
(229,52.70)
(230,52.70)
(231,52.70)
(232,52.70)
(233,52.70)
(234,52.70)
(235,52.70)
(236,52.70)
(237,52.70)
(238,52.70)
(239,52.70)
(240,52.70)
(241,52.70)
(242,52.70)
(243,52.70)
(244,52.70)
(245,52.70)
(246,52.70)
(247,52.70)
(248,52.70)
(249,52.70)
(250,52.70)
(251,52.70)
(252,52.70)
(253,52.70)
(254,52.70)
(255,52.70)
(256,52.70)
(257,52.70)
(258,52.70)
(259,52.70)
(260,52.70)
(261,52.70)
(262,52.70)
(263,52.70)
(264,52.70)
(265,52.70)
(266,53.43)
(267,53.43)
(268,53.43)
(269,53.43)
(270,53.43)
(271,53.43)
(272,53.43)
(273,53.43)
(274,53.43)
(275,53.43)
};

\addplot[
    mark=diamond*,                
    mark size=2pt,
    only marks,
    color=blue!70!black,
    thick
] coordinates {(265,53.43)};


\addplot[densely dashed,blue!70!black,thick] coordinates {(265,10) (265,53.43)};

\addplot[
    mark=diamond*,                
    mark size=2pt,
    only marks,
    color=blue!70!black,
    thick
] coordinates {(1,47.68)};


\addplot[densely dashed,blue!70!black,thick] coordinates {(1,10) (1,47.68)};

\addplot[dashed,black!60,thick,domain=0:275]{43.33};
\node[anchor=west,font=\scriptsize\bfseries,text=black]
    at (axis cs:275,43.43){LIME};
    

\addplot[dashed,black!60,thick,domain=0:275]{37.58};
\node[anchor=west,font=\scriptsize\bfseries,text=black]
    at (axis cs:275,38.35){Hawkeye};

\addplot[dashed,black!60,thick,domain=0:275]{37.56};
\node[anchor=west,font=\scriptsize\bfseries,text=black]
    at (axis cs:275,35.15){SHiP++};

\addplot[dashed,black!60,thick,domain=0:275]{34.66};
\node[anchor=west,font=\scriptsize\bfseries,text=black]
    at (axis cs:275,32.16){RCR};

\addplot[dashed,black!60,thick,domain=0:275]{11.45};
\node[anchor=west,font=\scriptsize\bfseries,text=black]
    at (axis cs:275,11.45){MPPPB};

\end{axis}
\end{tikzpicture}
\caption{Cache Hit Rate improvement using 4o-mini+Gemini 2.5 pro. The first vertical line shows the first instance where \textsc{CacheForge} surpasses a state-of-the-art policy and the second vertical line signifies the model change from 4o-mini to Gemini 2.5 pro.} 
\label{fig:hit_evo}
\end{figure}

\noindent\textbf{Hit Rate} Figure~\ref{fig:hit_evo} shows that average Hit rate continuously evolving over 260 iterations of \textsc{CacheForge} starts from 47.68\% and goes up to 53.43\%, a 4.67× times improvement over MPPPB. 
This number is 42.18\%, 54.15\%, 42.26\%, 23.31\%, and 28.16\% over Hawkeye, ReD, SHiP++, LIME, and LRU respectively.
A small but important detail is that progress plateaus near iteration 250; replacing the underlying LLM (from \texttt{o4-mini} to Gemini~2.5 Pro) at iteration 265 immediately lifts the hit rate from 52.70\% to 53.43\%, indicating that the late-stage plateau was model-limited rather than search-limited. 


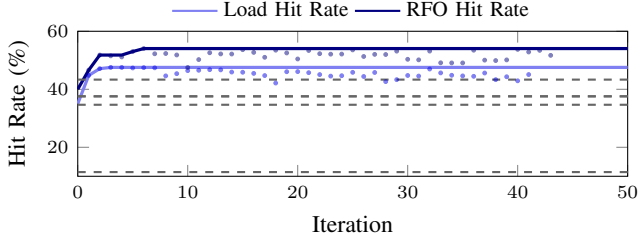
\begin{figure}[!htbp]
\centering
\begin{tikzpicture}
\begin{axis}[
    width=\columnwidth,
    height=3.5cm,
    xlabel={Iteration},
    ylabel={Hit Rate (\%)},
    xmin=0, xmax=50,
    ymin=10, ymax=60,
    legend style={
        font=\footnotesize,
        at={(0.5,1)}, 
        anchor=south,
        draw=none,
        fill=none,
    },
    legend columns=-1,
    tick label style={font=\scriptsize},
    label style={font=\small},
    legend cell align=left
]

\addlegendimage{line legend, blue!90!black!50, thick}
\addlegendentry{Load Hit Rate}
\addlegendimage{line legend, blue!50!black, thick}
\addlegendentry{RFO Hit Rate}

\addplot[blue!90!black!50, very thick, mark=none] coordinates {
(0,35.00)(1,44.79)(2,47.11)(3,47.54)(4,47.54)(5,47.54)(6,47.54)(7,47.54)
(8,47.54)(9,47.54)(10,47.54)(11,47.54)(12,47.54)(13,47.54)(14,47.54)
(15,47.54)(16,47.54)(17,47.54)(18,47.54)(19,47.54)(20,47.54)(21,47.54)
(22,47.54)(23,47.54)(24,47.54)(25,47.54)(26,47.54)(27,47.54)(28,47.54)
(29,47.54)(30,47.54)(31,47.54)(32,47.54)(33,47.54)(34,47.54)(35,47.54)
(36,47.54)(37,47.54)(38,47.54)(39,47.54)(40,47.54)(41,47.54)(45,47.54)
(50,47.54)
};

\addplot[only marks, mark=*, mark size=0.7pt, color=blue!90!black, opacity=0.5]
coordinates {
(1,44.79)(2,47.11)(3,47.54)(4,47.53)(5,47.35)(6,47.46)(7,47.40)(8,44.62)
(9,45.38)(10,46.43)(11,46.55)(12,46.76)(13,46.69)(14,45.94)(15,45.87)
(16,45.48)(17,44.73)(18,42.15)(19,46.05)(20,46.09)(21,45.73)(22,45.07)
(23,44.54)(24,44.68)(25,45.84)(26,44.41)(27,45.79)(28,42.64)(29,43.32)
(30,44.76)(31,44.56)(32,46.98)(33,45.65)(34,44.83)(35,44.62)(36,44.47)
(37,45.61)(38,43.32)(39,44.32)(40,42.92)(41,45.08)
};

\addplot[blue!50!black, very thick, mark=none] coordinates {
(0,40.00)(1,46.61)(2,51.80)(3,51.80)(4,51.80)(5,53.01)(6,54.04)(7,54.04)
(8,54.04)(9,54.04)(10,54.04)(11,54.04)(12,54.04)(13,54.04)(14,54.04)
(15,54.04)(16,54.04)(17,54.04)(18,54.04)(19,54.04)(20,54.04)(21,54.04)
(22,54.04)(23,54.04)(24,54.04)(25,54.04)(26,54.04)(27,54.04)(28,54.04)
(29,54.04)(30,54.04)(31,54.04)(32,54.04)(33,54.04)(34,54.04)(35,54.04)
(36,54.04)(37,54.04)(38,54.04)(39,54.04)(40,54.04)(41,54.04)(42,54.04)
(43,54.04)(50,54.04)
};

\addplot[only marks, mark=*, mark size=0.7pt, color=blue!50!black, opacity=0.5]
coordinates {
(1,46.61)(2,51.80)(3,51.52)(4,51.15)(5,53.01)(6,54.04)(7,52.24)(8,52.33)
(9,51.83)(10,47.52)(11,50.24)(12,52.59)(13,52.19)(14,52.16)(15,53.65)
(16,52.72)(17,51.03)(18,53.48)(19,52.53)(20,50.45)(21,53.37)(22,52.27)
(23,52.11)(24,52.14)(25,53.58)(26,51.31)(27,52.04)(28,51.99)(29,53.15)
(30,50.34)(31,50.15)(32,53.17)(33,49.13)(34,49.09)(35,49.09)(36,53.47)
(37,50.05)(38,49.90)(39,50.13)(40,53.76)(41,52.86)(42,53.42)(43,51.70)
};

\addplot[dashed,black!60,thick,domain=0:50]{43.33};
\addplot[dashed,black!60,thick,domain=0:50]{37.58};
\addplot[dashed,black!60,thick,domain=0:50]{37.56};
\addplot[dashed,black!60,thick,domain=0:50]{34.66};
\addplot[dashed,black!60,thick,domain=0:50]{11.45};

\end{axis}
\end{tikzpicture}
\caption{Comparison of Load and RFO Hit Rate evolution over iterations. Two blue tones denote the access types, with dashed lines indicating RFO progression. 
}
\label{fig:load_rfo_combined}
\end{figure}

Figure~\ref{fig:load_rfo_combined} shows how 
\textsc{CacheForge} progressively improves individual access-type hit rates across iterations. Breaking down the total hit rate reveals distinct microarchitectural trends. 
For example, \textit{Load hit rate}, represents demand fetch reuse in cache. \textsc{CacheForge} improves this type of hits as well, from 44.8\% to 47.5\% a 6.7\% improvement over the state-of-the-art baseline. 
\textit{RFO hit rate}, captures the locality of store ownership, which \textsc{CacheForge} enhances from 46.6\% to 54.0\%,  a 15.6\% improvement. 

\begin{figure}[!htbp]
\centering
\begin{tikzpicture}
\begin{axis}[
    width=\columnwidth,
    height=4.5cm,
    xlabel={Sorted Policies},
    ylabel={Cache Hit Rate (\%)},
    xmin=0, xmax=100,
    ymin=30, ymax=55,
    tick label style={font=\footnotesize},
    label style={font=\normalsize},
    every axis plot/.append style={mark=none,line width=1pt},
    legend columns=3,
    legend style={
        font=\footnotesize,
        at={(0.5,1)}, 
        anchor=south,
        draw=none,
        fill=none,
    },
]

\addplot[blue!100!black!40,very thick]
coordinates {
(1,50.32)
(2,50.29)
(3,50.22)
(4,50.20)
(5,50.20)
(6,50.18)
(7,50.14)
(8,50.04)
(9,49.99)
(10,49.94)
(11,49.89)
(12,49.71)
(13,49.60)
(14,49.58)
(15,49.53)
(16,49.49)
(17,49.48)
(18,49.34)
(19,49.19)
(20,49.14)
(21,49.14)
(22,49.13)
(23,49.13)
(24,49.12)
(25,49.11)
(26,49.11)
(27,49.05)
(28,49.03)
(29,49.03)
(30,48.99)
(31,48.96)
(32,48.89)
(33,48.84)
(34,48.77)
(35,48.74)
(36,48.64)
(37,48.52)
(38,48.51)
(39,48.48)
(40,48.40)
(41,48.39)
(42,48.38)
(43,48.30)
(44,48.22)
(45,48.14)
(46,48.12)
(47,47.84)
(48,47.72)
(49,47.72)
(50,47.23)
(51,47.21)
(52,47.21)
(53,47.20)
(54,47.16)
(55,47.12)
(56,47.11)
(57,47.10)
(58,47.06)
(59,46.94)
(60,46.89)
(61,46.87)
(62,46.80)
(63,46.71)
(64,46.65)
(65,46.61)
(66,46.57)
(67,46.57)
(68,46.53)
(69,46.52)
(70,46.45)
(71,46.44)
(72,46.38)
(73,46.32)
(74,46.22)
(75,45.91)
(76,45.82)
(77,45.78)
(78,45.65)
(79,45.49)
(80,45.23)
(81,45.22)
(82,45.07)
(83,44.94)
(84,44.78)
(85,44.77)
(86,44.68)
(87,44.58)
(88,44.49)
(89,44.42)
(90,43.83)
(91,43.69)
(92,43.48)
(93,43.40)
(94,42.99)
(95,42.47)
(96,42.38)
(97,40.88)
(98,40.16)
(99,36.35)
};
\addlegendentry{Weighted Average}

\addplot[blue!90!black!70,very thick]
coordinates {
(1,49.68)
(2,49.67)
(3,49.67)
(4,49.65)
(5,49.65)
(6,49.61)
(7,49.60)
(8,48.42)
(9,48.30)
(10,48.27)
(11,48.25)
(12,48.14)
(13,48.14)
(14,48.14)
(15,48.13)
(16,48.10)
(17,48.03)
(18,47.81)
(19,47.79)
(20,47.73)
(21,47.41)
(22,47.41)
(23,47.17)
(24,47.15)
(25,47.15)
(26,47.14)
(27,47.11)
(28,47.00)
(29,46.99)
(30,46.91)
(31,46.89)
(32,46.85)
(33,46.85)
(34,46.82)
(35,46.72)
(36,46.34)
(37,46.27)
(38,46.21)
(39,46.14)
(40,46.11)
(41,45.98)
(42,45.97)
(43,45.91)
(44,45.89)
(45,45.74)
(46,45.47)
(47,45.43)
(48,45.40)
(49,45.37)
(50,45.31)
(51,45.19)
(52,45.04)
(53,44.88)
(54,44.85)
(55,44.84)
(56,44.81)
(57,44.79)
(58,44.64)
(59,44.34)
(60,43.61)
(61,43.55)
(62,43.47)
(63,42.93)
(64,42.88)
(65,40.84)
(66,38.72)
(67,38.38)
(68,36.88)
};
\addlegendentry{Hit Rate}

\addplot[blue!50!black,very thick]
coordinates {
(1,49.54)
(2,49.51)
(3,49.12)
(4,48.87)
(5,48.82)
(6,48.40)
(7,48.35)
(8,48.29)
(9,48.23)
(10,48.04)
(11,47.71)
(12,47.65)
(13,47.63)
(14,47.60)
(15,47.58)
(16,47.13)
(17,46.80)
(18,46.59)
(19,46.51)
(20,46.10)
(21,46.05)
(22,46.04)
(23,45.69)
(24,45.69)
(25,45.55)
(26,45.32)
(27,44.69)
(28,44.16)
(29,44.09)
(30,43.67)
(31,43.31)
(32,43.02)
(33,42.82)
(34,42.77)
(35,42.73)
(36,42.56)
(37,42.51)
(38,42.27)
(39,42.07)
(40,42.02)
(41,41.56)
(42,41.54)
(43,41.53)
(44,41.47)
(45,41.45)
(46,41.00)
(47,40.82)
(48,40.59)
(49,40.29)
(50,40.25)
(51,40.10)
(52,40.01)
(53,40.00)
(54,39.86)
(55,39.84)
(56,39.83)
(57,39.76)
(58,39.37)
(59,39.34)
(60,39.15)
(61,38.77)
(62,38.67)
(63,38.58)
(64,38.40)
(65,38.28)
(66,38.13)
(67,37.98)
(68,37.70)
(69,37.59)
(70,37.37)
(71,37.22)
(72,36.98)
(73,36.97)
(74,36.95)
(75,36.73)
(76,36.73)
(77,36.66)
(78,36.53)
(79,36.49)
(80,36.26)
(81,35.97)
(82,35.94)
(83,35.77)
(84,35.68)
(85,35.67)
(86,35.65)
(87,35.64)
(88,35.54)
(89,35.53)
(90,32.07)
};
\addlegendentry{Average}

\addplot[dashed,black!60,thick,domain=0:100]{43.33};
\node[anchor=west,font=\tiny\bfseries,text=black] at (axis cs:100,43.33){LIME};

\addplot[dashed,black!60,thick,domain=0:100]{41.69};
\node[anchor=west,font=\tiny\bfseries,text=black] at (axis cs:100,41.69){LRU};

\addplot[dashed,black!60,thick,domain=0:100]{37.58};
\node[anchor=west,font=\tiny\bfseries,text=black] at (axis cs:100,38){Hawkeye};

\addplot[dashed,black!60,thick,domain=0:100]{37.56};
\node[anchor=west,font=\tiny\bfseries,text=black] at (axis cs:100,37){SHiP++};

\addplot[dashed,black!60,thick,domain=0:100]{34.66};
\node[anchor=west,font=\tiny\bfseries,text=black] at (axis cs:100,34.66){ReD};

\end{axis}
\end{tikzpicture}

\vspace{1em}

\centering
\begin{tikzpicture}
\begin{axis}[
    width=\columnwidth,
    height=4.5cm,
    xlabel={Sorted Policies},
    ylabel={IPC},
    xmin=0, xmax=100,
    ymin=0.3, ymax=0.335,
    tick label style={font=\footnotesize},
    label style={font=\normalsize},
    every axis plot/.append style={mark=none,line width=1pt},
    legend columns=3,
    legend style={
        font=\footnotesize,
        at={(0.5,1)}, 
        anchor=south,
        draw=none,
        fill=none,
    },
]

\addplot[blue!90!black!70,very thick]
coordinates {
(1,0.3324)
(2,0.3323)
(3,0.3320)
(4,0.3312)
(5,0.3312)
(6,0.3312)
(7,0.3311)
(8,0.3311)
(9,0.3310)
(10,0.3307)
(11,0.3307)
(12,0.3306)
(13,0.3305)
(14,0.3303)
(15,0.3303)
(16,0.3299)
(17,0.3294)
(18,0.3289)
(19,0.3288)
(20,0.3281)
(21,0.3279)
(22,0.3279)
(23,0.3278)
(24,0.3277)
(25,0.3276)
(26,0.3265)
(27,0.3248)
(28,0.3244)
(29,0.3240)
(30,0.3240)
(31,0.3240)
(32,0.3239)
(33,0.3239)
(34,0.3236)
(35,0.3234)
(36,0.3230)
(37,0.3229)
(38,0.3228)
(39,0.3227)
(40,0.3225)
(41,0.3223)
(42,0.3220)
(43,0.3220)
(44,0.3219)
(45,0.3218)
(46,0.3217)
(47,0.3217)
(48,0.3214)
(49,0.3210)
(50,0.3208)
(51,0.3204)
(52,0.3202)
(53,0.3200)
(54,0.3199)
(55,0.3194)
(56,0.3185)
(57,0.3178)
(58,0.3177)
(59,0.3177)
(60,0.3177)
(61,0.3172)
(62,0.3171)
(63,0.3162)
(64,0.3154)
(65,0.3154)
(66,0.3152)
(67,0.3139)
(68,0.3138)
(69,0.3130)
(70,0.3129)
(71,0.3127)
(72,0.3124)
(73,0.3122)
(74,0.3120)
(75,0.3117)
(76,0.3117)
(77,0.3113)
(78,0.3110)
(79,0.3110)
(80,0.3109)
(81,0.3109)
(82,0.3106)
(83,0.3105)
(84,0.3105)
(85,0.3104)
(86,0.3102)
(87,0.3095)
(88,0.3092)
(89,0.3091)
(90,0.3091)
(91,0.3090)
(92,0.3087)
(93,0.3084)
(94,0.3065)
(95,0.3061)
(96,0.3038)
(97,0.3037)
};
\addlegendentry{IPC}

\addplot[blue!50!black,very thick]
coordinates {
(1,0.3333)
(2,0.3329)
(3,0.3325)
(4,0.3320)
(5,0.3319)
(6,0.3319)
(7,0.3317)
(8,0.3316)
(9,0.3316)
(10,0.3314)
(11,0.3313)
(12,0.3312)
(13,0.3312)
(14,0.3312)
(15,0.3310)
(16,0.3309)
(17,0.3308)
(18,0.3307)
(19,0.3307)
(20,0.3307)
(21,0.3307)
(22,0.3307)
(23,0.3306)
(24,0.3303)
(25,0.3302)
(26,0.3298)
(27,0.3294)
(28,0.3294)
(29,0.3280)
(30,0.3278)
(31,0.3278)
(32,0.3278)
(33,0.3278)
(34,0.3276)
(35,0.3276)
(36,0.3263)
(37,0.3244)
(38,0.3243)
(39,0.3239)
(40,0.3238)
(41,0.3233)
(42,0.3229)
(43,0.3227)
(44,0.3225)
(45,0.3221)
(46,0.3220)
(47,0.3211)
(48,0.3210)
(49,0.3205)
(50,0.3186)
(51,0.3185)
(52,0.3167)
(53,0.3166)
(54,0.3166)
(55,0.3163)
(56,0.3159)
(57,0.3157)
(58,0.3154)
(59,0.3149)
(60,0.3143)
(61,0.3142)
(62,0.3139)
(63,0.3139)
(64,0.3138)
(65,0.3136)
(66,0.3136)
(67,0.3132)
(68,0.3132)
(69,0.3129)
(70,0.3128)
(71,0.3123)
(72,0.3122)
(73,0.3121)
(74,0.3116)
(75,0.3115)
(76,0.3109)
(77,0.3108)
(78,0.3107)
(79,0.3095)
(80,0.3080)
(81,0.3073)
(82,0.3057)
(83,0.3045)
(84,0.3033)
(85,0.3024)
(86,0.3017)
(87,0.3016)
(88,0.3014)
(89,0.3014)
(90,0.3012)
};
\addlegendentry{Average}






\addplot[dashed,black!60,thick,domain=0:100]{0.3032};
\node[anchor=west,font=\scriptsize\bfseries,text=black]
    at (axis cs:100,0.3032){LRU};

\addplot[dashed,black!60,thick,domain=0:100]{0.3096};
\node[anchor=west,font=\scriptsize\bfseries,text=black]
    at (axis cs:100,0.3096){MPPPB};

\addplot[dashed,black!60,thick,domain=0:100]{0.3141};
\node[anchor=west,font=\scriptsize\bfseries,text=black]
    at (axis cs:100,0.3141){LIME};

\addplot[dashed,black!60,thick,domain=0:100]{0.3223};
\node[anchor=west,font=\scriptsize\bfseries,text=black]
    at (axis cs:100,0.3223){RCR};

\addplot[dashed,black!60,thick,domain=0:100]{0.3239};
\node[anchor=west,font=\scriptsize\bfseries,text=black]
    at (axis cs:100,0.3239){SHiP++};

\addplot[dashed,black!60,thick,domain=0:100]{0.3252};
\node[anchor=west,font=\scriptsize\bfseries,text=black]
    at (axis cs:100,0.3252){Hawkeye};

\end{axis}
\end{tikzpicture}

\caption{Sorted cache hit rates for policies evolved under Hit Rate, Average, and Weighted Average rewards and sorted IPC for policies evolved under IPC, and Average rewards.
} 
\label{fig:reward_sorted_onecol}
\end{figure}
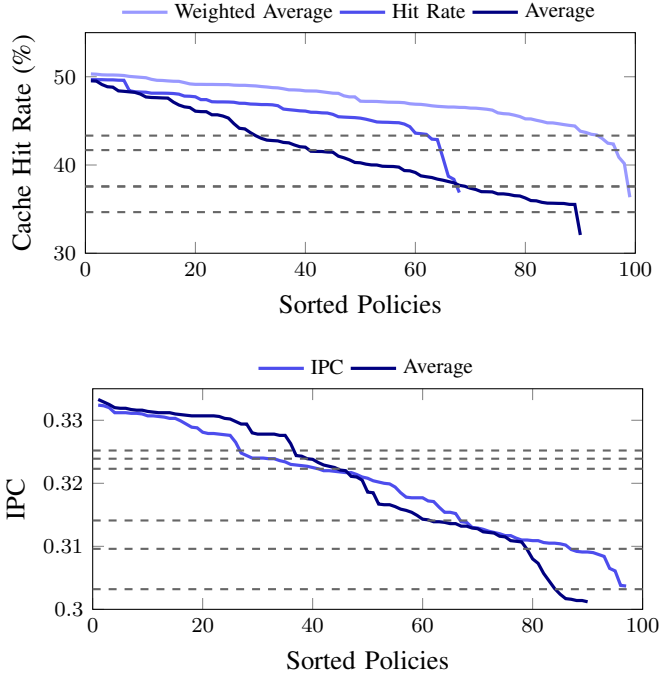


\noindent\textbf{Weighted Reward Function.} 
Each generated policy is scored using a composite reward function that aligns the simulator’s raw statistics with architectural performance goals. 
We evaluate three formulations: (\textit{i}) a Hit-Rate Reward emphasizing overall cache reuse, (\textit{ii}) an Average Reward combining hit rate and IPC equally, and (\textit{iii}) a Weighted Average Reward that biases towards IPC rather than hit rate. 
Empirically, this design yields the best performing policies, improving total hit rate by 3–5\% over CacheForge with IPC alone as the reward and demonstrating that reward shaping provides a powerful means of guiding LLM-driven optimization toward high-value microarchitectural trade-offs (see Figure~\ref{fig:reward_sorted_onecol}). 

\noindent\textbf{Generalization vs. Overfitting evaluation.} 
To measure the generalization capability of \textsc{CacheForge}, we autonomously synthesized policies over 10 M-instruction window using exploration phase for faster evolution and \textit{five} training workloads, and evaluated them across all 28 SPEC workloads for 1B instruction. On average, it achieves a 15.52\%  (5.73\%, 11.83\%, 27.36\%, 13.15\%, 13.72\%, 19.69\% over LRU, Less Is More, MPPPB,  SHiP++,  Hawkeye, and ReD) 
improvement in total hit rate over the baseline policies. 


Taken together, these results show that \textsc{CacheForge} 
discovers, through trial and validation on the real simulator, \emph{what the metric should have been in the first place}. 
By observing thousands of candidates, \textsc{CacheForge} learns that total hit rate is at best incidental, and at worst misleading, and gradually shifts its behavior toward improving the access types and structural patterns that genuinely raise IPC. 

Once the reward is aligned with these causal signals, the system consistently refines policies toward higher Load and RFO hit rates, shorter exposed miss latencies, and better IPC-per-KB efficiency without any offline training or human heuristics.

\begin{figure}[!htbp]
    \centering
    \includegraphics[width=1\linewidth, height=4cm]{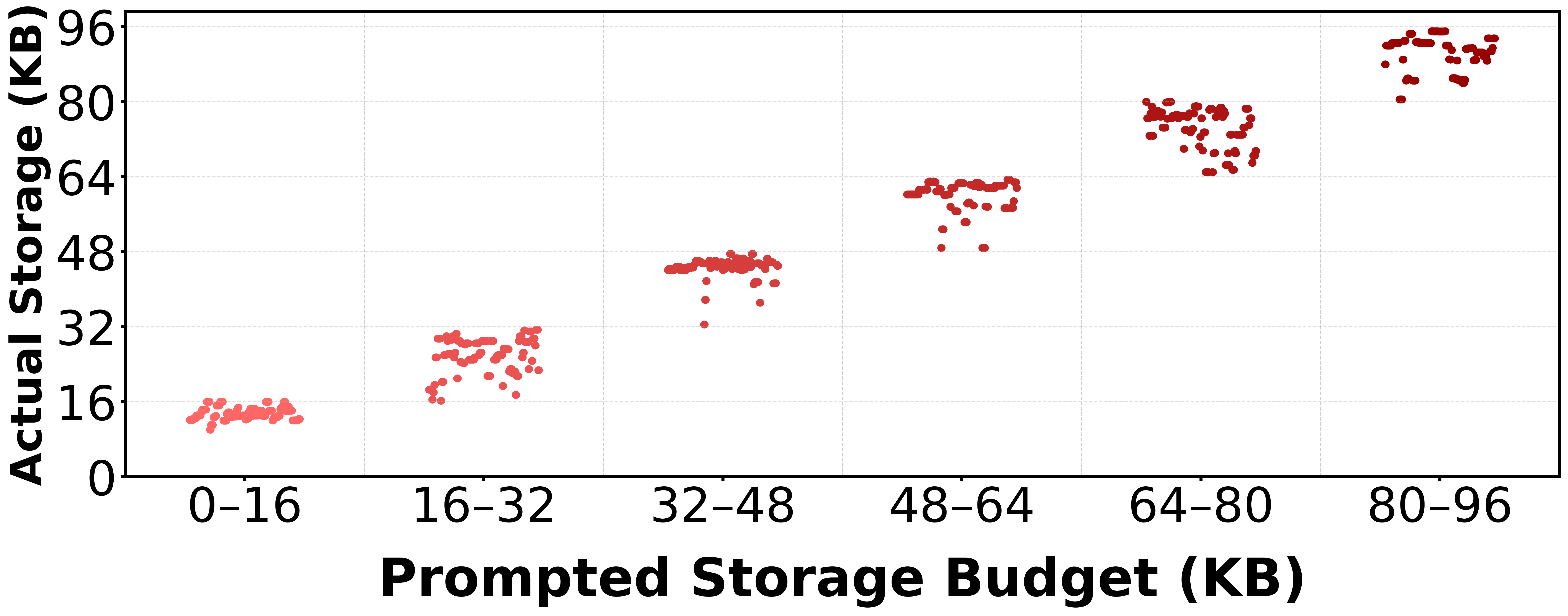}
    \caption{Successful replacement policy hardware overhead budget  alignment} 
    \label{fig:storage_alignment}
\end{figure}

\begin{figure}[!htbp]
\begin{tikzpicture}
    \begin{axis}[
        ybar,
        bar width=12pt,
        width=9cm,
        height=3.5cm,
        ymin=0,
        ylabel={Storage (KB)},
        symbolic x coords={
        Hawkeye, ReD, LIME, MPPPB, SHiP++, LRU, CacheForge},
        xtick=data,
        x tick label style={rotate=45, anchor=east},
        enlarge x limits=0.15
    ]
    \addplot[fill=black] coordinates {
        (LRU,16)
        (Hawkeye,31.80)
        (ReD,31.875)
        (SHiP++, 20)
        (LIME,31.20)
        (MPPPB,29.05)
        (CacheForge,12)
    };
    \end{axis}
\end{tikzpicture}
    \caption{Storage Overhead (KB) of CacheForge Proposed Policies vs. Baseline Policies }
    \label{fig:Storage_Overhead}
\end{figure}
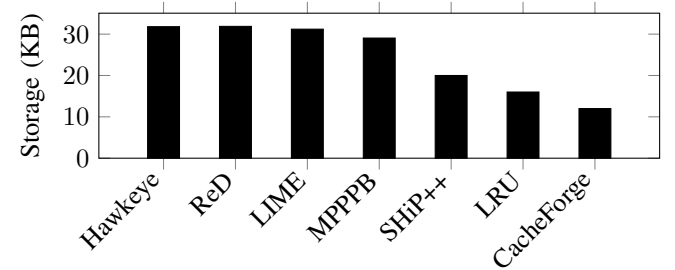

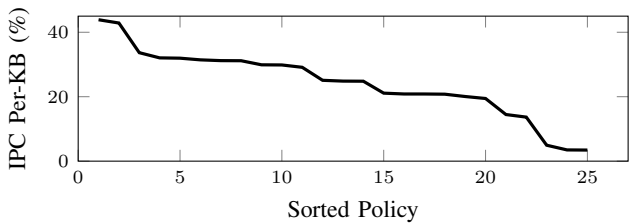
\begin{figure}[!htbp]
\centering
\begin{tikzpicture}
\begin{axis}[
    width=\columnwidth,
    height=3.5cm,
    xlabel={Sorted Policy},
    ylabel={IPC Per-KB (\%)},
    xmin=0, xmax=27,
    ymin=0, ymax=45,
    tick label style={font=\scriptsize},
    label style={font=\small},
    every axis plot/.append style={mark=none,line width=0.9pt},
    clip=false,
    legend style={
        font=\footnotesize,
        at={(0.02,0.98)},
        anchor=north west,
        draw=none,
        fill=none,
        legend cell align=left,
    },
]

\addplot[black,very thick]
coordinates {
(1,43.882239)
(2,42.863373)
(3,33.659899)
(4,32.048966)
(5,31.954505)
(6,31.434814)
(7,31.194731)
(8,31.164114)
(9,29.898013)
(10,29.850896)
(11,29.121297)
(12,25.087427)
(13,24.839142)
(14,24.806469)
(15,21.101234)
(16,20.843707)
(17,20.840920)
(18,20.779193)
(19,20.058940)
(20,19.442737)
(21,14.467586)
(22,13.663531)
(23,4.944471)
(24,3.485242)
(25,3.441947)
};

\end{axis}
\end{tikzpicture}
\caption{IPC per KB improvement over Multiperspective.} 
\label{fig:mpppb_ipc_per_kb_improvement}
\end{figure}

\subsection{Evolving at Controlled and Low Budget}
\noindent\textbf{Hardware Overhead. } 
Figure~\ref{fig:storage_alignment} shows that \textsc{CacheForge} reliably
follows storage constraints. When prompted with different storage-budget
ranges, the generated policies cluster tightly within the corresponding
actual storage bands, indicating that the LLM not only understands the
budget specification but produces implementations that remain aligned
with it. This confirms that storage limits can be enforced effectively
through prompting alone in CacheForge.

Figure~\ref{fig:Storage_Overhead} presents the storage overhead of the proposed \textsc{CacheForge} policy relative to established CRC-2 baselines. The baseline policies require between 16\,KB and 31.875\,KB of LLC metadata: LRU (16\,KB), SHiP++ (20\,KB), MPPPB (29.05\,KB), LIME (31.20\,KB), Hawkeye (31.80\,KB), and ReD (31.875\,KB). In contrast, \textsc{CacheForge} uses only 12\,KB. This represents a substantial reduction, \textsc{CacheForge} requires 43\% less storage than MPPPB, the most comparable high-performing baseline, while still delivering higher IPC/KB efficiency (Figure~\ref{fig:mpppb_ipc_per_kb_improvement}).

\input{Results/Optimization}

\input{Results/IPC}

\input{Results/hardware}
\input{Results/speedup}

\begin{table*}[!ht]
\centering
\setlength{\tabcolsep}{4pt}

\begin{tabularx}{\textwidth}{
    >{\raggedright\arraybackslash}p{0.2\textwidth}   
    >{\raggedright\arraybackslash}p{0.23\textwidth}   
    >{\raggedright\arraybackslash}p{0.23\textwidth}   
    >{\raggedright\arraybackslash}p{0.34\textwidth}          
}
\toprule
\textbf{Novel Design Component} &
\textbf{Objective} &
\textbf{Impact} &
\textbf{CacheForge Capability} \\
\midrule

Reinforcement-Learning & 
Learn eviction value online &
Qualitative only &
Explores unconventional algorithmic spaces \\

Q-Value Candidate Ranking &
Score all candidates &
High granularity eviction decisions &
Applies ranking-based eviction strategies \\

Online Reward Adaptation &
Adjust RL policy using hits/misses &
Adaptive learning &
Experiments with reinforcement learning \\


Dead-Block Prediction &
Identify low-reuse lines &
Improves reuse filtering &
Rediscovers classical dead block mechanisms \\

Evicted-Address Filter &
Detect recently-dead lines &
Reduces pollution &
Uses compact deadness filters effectively \\

Prefetch Usefulness &
Reduce unused prefetches &
Reduces cache pollution &
Models prefetch awareness \\

Region Hotness Tracking &
Identify hot or thrashing regions &
Reduces repeated thrashing &
Infers region-level memory behavior \\

Ghost Re-Reference Buffer &
Detect premature evictions &
Corrects eviction mistakes &
 ghost-based structures \\

Probabilistic Stack-Distance &
Estimate reuse using recency stack &
Improves graded reuse prediction &
Investigates probabilistic reuse modeling \\

MRU Promotion Control &
Limit noisy MRU promotions &
Reduces pollution due to bursts &
understand promotion strategies\\

Adaptive Bypass Control &
Avoid caching low reuse lines &
Reduces pollution &
Understands bypass rules \\

Stride Detection &
Detect stride accesses &
Improves predictable performance &
Recognizes regular temporal locality \\

Stream Detection &
Detect streaming behavior &
Reduces streaming pollution &
Identifies spatial locality \\

PC Friendliness Scoring &
Differentiate friendly vs.\ averse PCs &
Improves context-aware reuse &
Learns PC/context-driven locality \\

Global Predictor Decay &
Prevent counter saturation &
Maintains long-term stability &
Manages dynamic predictor state \\

Weighted Eviction Ranking &
Score MAX-RRPV candidates &
More precise victim selection &
Constructs multi-factor scoring methods \\

\bottomrule
\end{tabularx}
\caption{Novel Design Components explored by CacheForge.}
\label{tab:novel_designs}
\end{table*}

\begin{figure*}[!htpb]
    \centering
    \includegraphics[width=\linewidth]{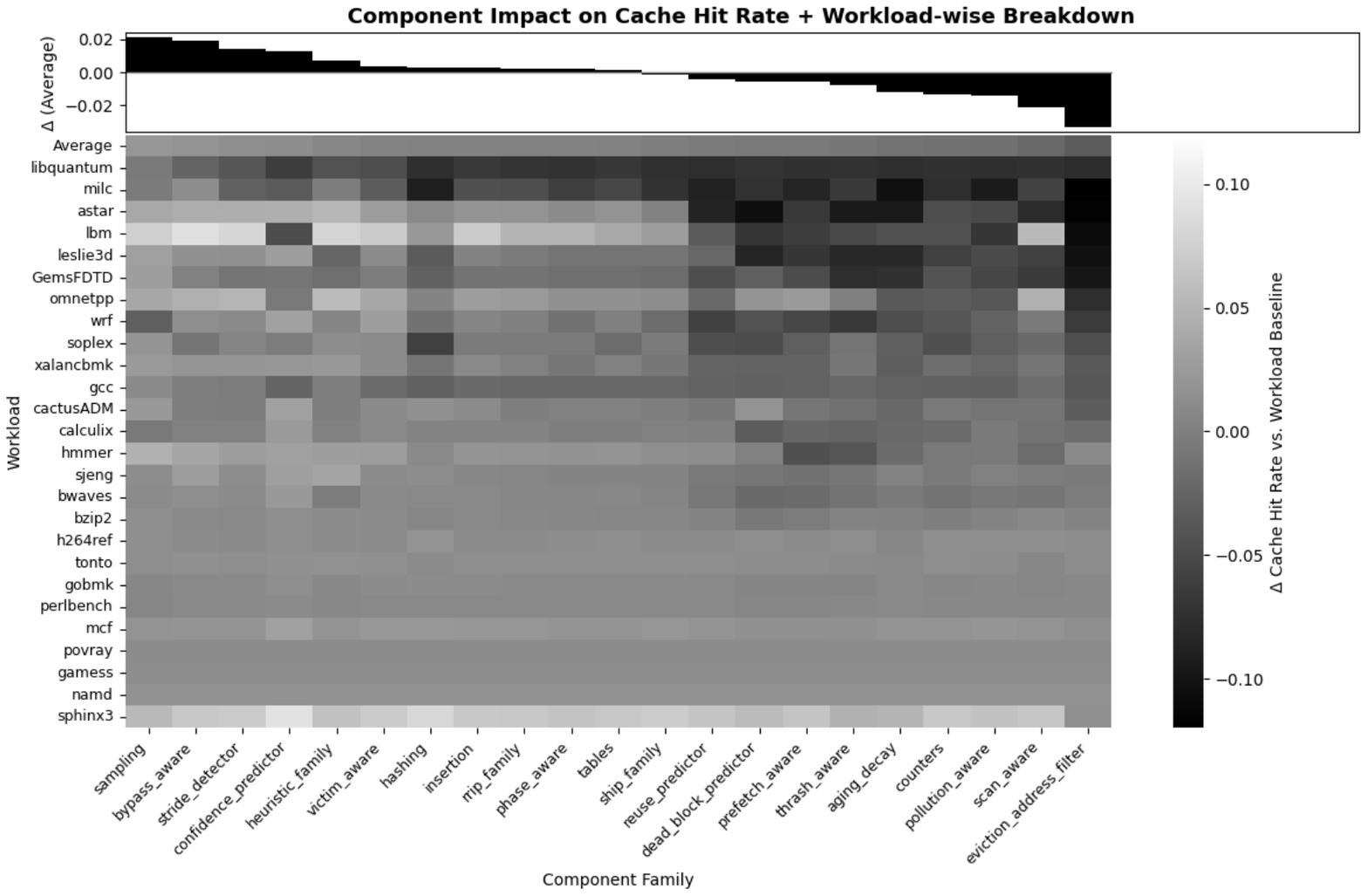} 
    \caption{Heatmap showing the impact of novel features explored in CacheForge over Hit Rate.}
    \label{fig:heatmap}
\end{figure*}



\section{Novel \textsc{CacheForge} Policies.}

\begin{table*}[!htbp]
\centering
\setlength{\tabcolsep}{3pt}

\begin{tabular}{
  p{2cm}    
  p{5cm}      
  *{5}{p{0.55cm}}   
  *{5}{p{0.55cm}}   
  p{1.8cm}    
}
\toprule
\textbf{Component} & \textbf{Description} &
\multicolumn{5}{c}{\textbf{Multiperspective}} &
\multicolumn{5}{c}{\textbf{CacheForge}} &
\textbf{Tuning} \\

\cmidrule(lr){3-7} \cmidrule(lr){8-12}

& & A & B & E & W & X & A & B & E & W & X & \\
\midrule


\multicolumn{13}{l}{\textbf{Perceptron Features}} \\
\midrule
F\_PC &
\multirow{7}{5cm}{PC: Bits B-E of the PC of the $W$th memory access instruction.
A is the recency-stack position beyond which the block is considered dead.} 
& 7 & 14 & 43 & 11 & 0 & 6 & 0 & 16 & 8 & 0 & Changed \\

F\_PC & & 16 & 3 & 11 & 16 & 1 & 8 & 6 & 22 & 3 & 0 & Changed \\
F\_PC & & 10 & 1 & 53 & 10 & 0 & 10 & 4 & 20 & 1 & 0 & Changed \\
F\_PC & & 16 & 8 & 16 & 5 & 0 & 12 & 4 & 20 & 0 & 2 & Changed \\
F\_PC & & 17 & 6 & 20 & 0 & 1 & -- & -- & -- & -- & -- & Removed \\
F\_PC & & 17 & 6 & 20 & 0 & 1 & -- & -- & -- & -- & -- & Removed \\
F\_PC & & 17 & 6 & 20 & 14 & 1 & -- & -- & -- & -- & -- & Removed \\

\midrule

F\_OFF &
\multirow{3}{5cm}{Offset bits (B-E) extracted from the block offset of memory access.} 
& 15 & 1 & 6 & 0 & 1 & 8 & 4 & 6 & 0 & 0 & Changed \\

F\_OFF & & 10 & 0 & 6 & 0 & 1 & 10 & 2 & 6 & 0 & 0 & Changed \\
F\_OFF & & -- & -- & -- & -- & -- & 14 & 0 & 6 & 0 & 0 & Added \\

\midrule

F\_LM &
Lastmiss bit indicating if last set access missed.
& 9 & 0 & 0 & 0 & 0 & 10 & 2 & 6 & 0 & 0 & Changed \\

\midrule

F\_INS &
\multirow{4}{5cm}{Insertion bit indicating whether the access is marked as an insertion.} 
& 16 & 0 & 0 & 0 & 1 & 8 & 0 & 0 & 0 & 0 & Changed \\

F\_INS & & 8 & 0 & 0 & 0 & 1 & -- & -- & -- & -- & -- & Removed \\
F\_INS & & 16 & 2 & 47 & 2 & 0 & -- & -- & -- & -- & -- & Removed \\
F\_INS & & 17 & 0 & 0 & 0 & 1 & -- & -- & -- & -- & -- & Removed \\

\midrule

F\_BURST & Burst bit indicating cache burst. &
 6 & 11 & 22 & 9 & 0 & 6 & 0 & 0 & 0 & 2 & Changed \\

\midrule

F\_BIAS & Bias bit. &
16 & 0 & 0 & 0 & 0 & 16 & 0 & 0 & 0 & 0 & No Change \\

\midrule

F\_TAG &
\multirow{3}{5cm}{Physical address bits (B-E) extracted from the address.} 
& -- & -- & -- & -- & -- & 12 & 30 & 42 & 0 & 0 & Added \\

F\_TAG & & -- & -- & -- & -- & -- & 16 & 12 & 21 & 0 & 0 & Added \\
F\_TAG & & -- & -- & -- & -- & -- & 16 & 18 & 36 & 0 & 0 & Added \\


\midrule
\multicolumn{13}{l}{\textbf{Sampler}} \\
\midrule
Number & Number of samplers &
80 & & & & & \textbf{48} & & & & & Changed \\

Associativity & Associativity of samplers &
18 & & & & & \textbf{12} & & & & & Changed \\

Access Type & Logic to differentiate access types &
\multicolumn{5}{l}{Prefetch + Writeback} &
\multicolumn{5}{l}{RFO + Prefetch + Writeback} &
Changed \\


\midrule
\multicolumn{13}{l}{\textbf{Placement Vector Thresholds}} \\
\midrule
plv[i][j] & Placement Level Vector values &
\multicolumn{5}{l}{$-15,0,35,12,44,15$} &
\multicolumn{5}{l}{$-8,0,52,9,112,15$} &
Changed \\


\midrule
\multicolumn{13}{l}{\textbf{Implementation Changes and Confidence Values}} \\
\midrule
Confidence Thresholds & Confidence values (dt2, theta, theta\_2, promotion threshold). &
\multicolumn{5}{l}{256, 109, 135, 82} &
\multicolumn{5}{l}{176, 96, 152, 112} &
Changed \\

Predictor Tables & Number of predictor tables. &
\multicolumn{5}{l}{16 tables} &
\multicolumn{5}{l}{\textbf{14 tables}} &
Changed \\

Record Types & Types of accesses recorded. &
\multicolumn{5}{l}{Load, RFO, Prefetch, Writeback} &
\multicolumn{5}{l}{Load, RFO only} &
Changed \\

 Noise-controlled XOR & Applies mask on PC-bit slice &
\multicolumn{5}{l}{Does not exist} &
\multicolumn{5}{l}{XOR on partial PC slice} &
Novelty introduced \\

\bottomrule
\end{tabular}
\caption{Design comparison of CacheForge vs. Multiperspective.}
\label{tab:policy_comparison}
\end{table*}

Table~\ref{tab:novel_designs} details over some of the novel design components included in the top policies designed by CacheForge across various experiment configurations. 
The table also captures some of the existing and new ideas in cache replacement domain that are explored by CacheForge by combining some of these ideas into existing replacement policies to come up with novel policies. 

We analyzed 21 features  across 491 high performing CacheForge policies
and sorted them by impact on cache hit rate (Figure~\ref{fig:heatmap}). 
Our studies revealed that designs with certain features outperform others. For example, set sampling, bypassing, stride detectors, and confidence predictors have a higher average impact on cache hit rate than designs that had eviction address filters, scan awareness and pollution awareness components. Subsequently, we briefly analyze some of the top policies generated by CacheForge.

\noindent{\textbf{Spatial PC Aware Reuse Correlator+}}
\textbf{(SPARC+)}, the best IPC-oriented policy generated by CacheForge, improves IPC by  
10.15\% over LRU, across a subset of five high-MPKI, cache-sensitive SPEC workloads. 
SPARC+ adds a Graph Node Table to track spatial locality, reuse aware victim selection logic, hybrid insertion, and PC-based victim feedback to SHCT table.  
%
SPARC+
also applies a light periodic decay of graph score to avoid saturation and to maintain stable behavior over long traces. These combined refinements allow SPARC+ to improve reuse efficiency and latency performance, demonstrating CacheForge’s ability to produce simple yet effective enhancements to a well-established replacement policy.

\noindent{\textbf{Thrash-Aware Dual-Pattern Insertion with SHiP Reuse (TA-DPIP)}}, a compact policy 
is synthesized by CacheForge when it was reconfigured to optimize purely for total cache hit rate.
The policy retains the standard 2-bit RRIP victim selection mechanism and augments it with several lightweight, signature-indexed signals that modulate insertion priority: a SHiP-style temporal reuse counter, a stride-regularity counter that strengthens when the current line index repeats the last non-zero stride, a pointer-likeness counter that strengthens when the stride changes, and a small thrash-propensity counter that increases when lines from a signature are evicted without being referenced. These signals allow TA-DPIP to separate predicted-reuse signatures from streaming, pointer-heavy, or thrashing behaviors. 

TA-DPIP illustrates that CacheForge can synthesize policies that are both interpretable and hardware-efficient. 
\noindent\textbf{Region–StreamPath Multiperspective Predictor (RSP-MPPP)}, a novel policy synthesized by CacheForge, achieves IPC/KB improvement over MPPPB by 135\%, and LRU which has the best baseline IPC/KB by nearly 30\%. To achieve this performance, RSP-MPPP, underperforms MPPPB for IPC by 1.54\%, while cutting the storage requirement down by 49.37\% 
over MPPPB. On the other hand, it improves IPC over LRU by 1.08\% while 
reducing area by 21.97\% over LRU.


Table~\ref{tab:policy_comparison} details on the tuning and novel addition made by CacheForge to the baseline MPPPB policy. Collectively, these modifications reduce sampler and predictor overhead while stabilizing confidence estimation, yielding improved IPC/KB performance at a smaller footprint. Thus, CacheForge’s RSP-MPPP demonstrates the LLM’s ability to identify and tune critical thresholds, gating mechanisms, and feature structures within the Multiperspective framework, improving reuse prediction quality and cache efficiency under strict storage constraints.

\noindent\textbf{Set-Adaptive Gated Ensemble (SAGE-MPPPB)} 
While optimizing for IPC/KB across iterations, \textsc{CacheForge} also eventually improves the IPC performance
relative to MPPPB. Using MPPPB as the initial seed, the system ultimately discovers Set-Adaptive Gated Ensemble (SAGE-MPPPB), the first generated policy to surpass MPPPB. SAGE-MPPPB tunes MPPPB to bring down storage requirements by 29.53\% while adopting per-set hotness counting to achieve a 0.14\% speedup. SAGE-MPPPB effectively improves IPC/KB over Multiperspective by 43.88\%, demonstrating LLM's capability to improve performance while keeping storage budget constrained.
\input{Results/IPC-MPPPB-barchart}

\section{Sensitivity Studies}
\label{sec:sensitivity}

\begin{figure}[!htbp]
  \centering
  \includegraphics[width=\linewidth, height=3.5cm]{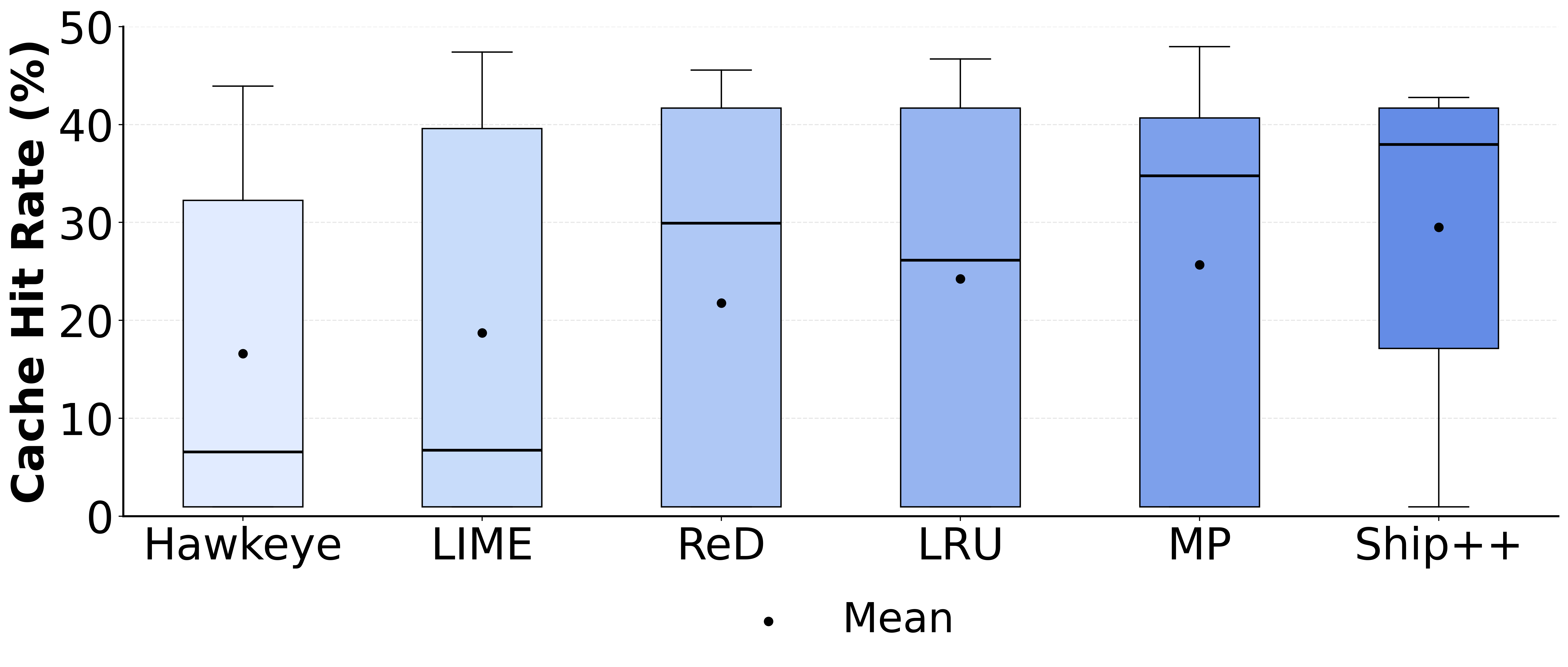}
  \caption{Seed analysis across various replacement policies.} 
  \label{fig:seed}
\end{figure}

\noindent{\textbf{Varying Seed Policy.}}  We further test experiments with and without an explicit \textit{seed policy}, in which the LLM either begins from an existing human-designed baseline 
or starts from a null design space. 
These seed and memory variations quantify the degree of dependency on prior knowledge and the model’s capacity for self-discovery. Figure~\ref{fig:seed}  presents a box plot comparing cache hit rate distributions across 6 policies: Hawkeye, LIME, ReD, LRU, MPPPB, and ShiP++
, revealing that ShiP++ achieved the highest average performance (29.5\%), followed by MPPPB (25.5\%). 
This pattern suggests that certain policy seeds, such as Multiperspective and Ship++, tend to generate more effective cache strategies overall, whereas Hawkeye-based seeds produced less efficient evolutionary results. 

\begin{figure}[!htbp]
\centering
\begin{tikzpicture}

\begin{axis}[
    width=8cm,
    height=4cm,
    xlabel={Iteration},
    ylabel={IPC},
    xmin=0, xmax=50,          
    ymin=0.3, ymax=0.3375,  
   y tick label style={
        /pgf/number format/fixed,       
        /pgf/number format/precision=3  
    },
    tick label style={font=\scriptsize},
    label style={font=\small},
    every axis plot/.append style={mark=none,line width=1pt},
    clip=false,
    legend style={
        font=\footnotesize,
        at={(0.02,0.98)}, 
        anchor=north west,
        draw=black,
        fill=white,
        legend cell align=left,
    },
    legend columns=-1,
]

\addplot[blue,dotted]
coordinates {
(1,0.30511)
(2,0.30631)
(3,0.30652)
(4,0.30652)
(5,0.30731)
(6,0.30731)
(7,0.30731)
(8,0.30731)
(9,0.30731)
(10,0.30731)
(11,0.30731)
(12,0.30731)
(13,0.31259)
(14,0.31407)
(15,0.31407)
(16,0.31535)
(17,0.31535)
(18,0.31535)
(19,0.31535)
(20,0.31535)
(21,0.31535)
(22,0.31535)
(23,0.31535)
(24,0.31535)
(25,0.31535)
(26,0.31535)
(27,0.31535)
(28,0.31535)
(29,0.31535)
(30,0.31535)
(31,0.31535)
(32,0.32109)
(33,0.32109)
(34,0.32109)
(35,0.32109)
(36,0.32109)
(37,0.32186)
(38,0.32186)
(39,0.32186)
(40,0.32186)
(41,0.32186)
(42,0.32186)
(43,0.32186)
(44,0.32684)
(45,0.32684)
(46,0.32684)
(47,0.32684)
(48,0.32684)
(49,0.32684)
(50,0.32684)
};
\addlegendentry{Best 2}

\addplot[green,densely dashed]
coordinates {
(1,0.30432)
(2,0.30908)
(3,0.30912)
(4,0.30912)
(5,0.30912)
(6,0.31753)
(7,0.31753)
(8,0.31753)
(9,0.31753)
(10,0.31753)
(11,0.31753)
(12,0.31753)
(13,0.31753)
(14,0.31753)
(15,0.31753)
(16,0.31753)
(17,0.31753)
(18,0.31753)
(19,0.31753)
(20,0.31753)
(21,0.31753)
(22,0.31753)
(23,0.31753)
(24,0.31753)
(25,0.31838)
(26,0.31838)
(27,0.31838)
(28,0.31838)
(29,0.31838)
(30,0.31838)
(31,0.31838)
(32,0.31838)
(33,0.31838)
(34,0.31838)
(35,0.31838)
(36,0.31838)
(37,0.31838)
(38,0.31838)
(39,0.31838)
(40,0.31838)
(41,0.31838)
(42,0.31838)
(43,0.31838)
(44,0.31838)
(45,0.31838)
(46,0.31838)
(47,0.32184)
(48,0.32184)
(49,0.32184)
(50,0.32184)
};
\addlegendentry{Best 5}

\addplot[red,very thick]
coordinates {
(1,0.30388)
(2,0.30437)
(3,0.30437)
(4,0.32085)
(5,0.32085)
(6,0.32085)
(7,0.32085)
(8,0.32085)
(9,0.32102)
(10,0.32102)
(11,0.32102)
(12,0.32102)
(13,0.32102)
(14,0.32102)
(15,0.32102)
(16,0.32102)
(17,0.32119)
(18,0.32121)
(19,0.32129)
(20,0.32129)
(21,0.32129)
(22,0.32129)
(23,0.32129)
(24,0.32129)
(25,0.32129)
(26,0.32129)
(27,0.32129)
(28,0.32129)
(29,0.32129)
(30,0.32129)
(31,0.32129)
(32,0.32129)
(33,0.32129)
(34,0.32276)
(35,0.32327)
(36,0.32327)
(37,0.32327)
(38,0.32327)
(39,0.32327)
(40,0.32327)
(41,0.32327)
(42,0.32365)
(43,0.32770)
(44,0.32770)
(45,0.32770)
(46,0.32770)
(47,0.32770)
(48,0.32770)
(49,0.32770)
(50,0.32782)

};
\addlegendentry{Best 10}

\end{axis}
\end{tikzpicture}
\caption{Impact of number of seeds to cross over}

\label{fig:seed_number}
\end{figure}
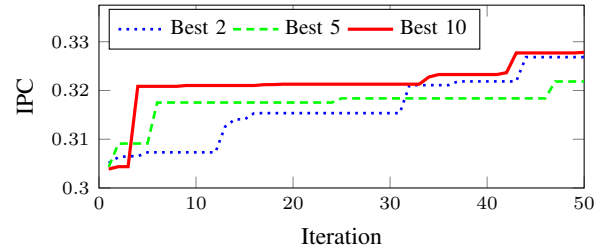

\noindent{\textbf{Varying Number of Seed Policy.}} We further assessed the impact of seed on IPC by varying to the number of input seeds shown in Figure~\ref{fig:seed_number}. We observed that the amount of context given to the LLM
influences its ability to refine prior designs, but not in a linear way: more examples do not always yield better results. The strongest
performance occurred when the LLM was given the top \textit{ten} policies, with
the top two performing second best; both outperformed the configuration using the top five. 

\begin{figure}[!htbp]
  \centering
  \includegraphics[width=1\linewidth, height=3.5cm]{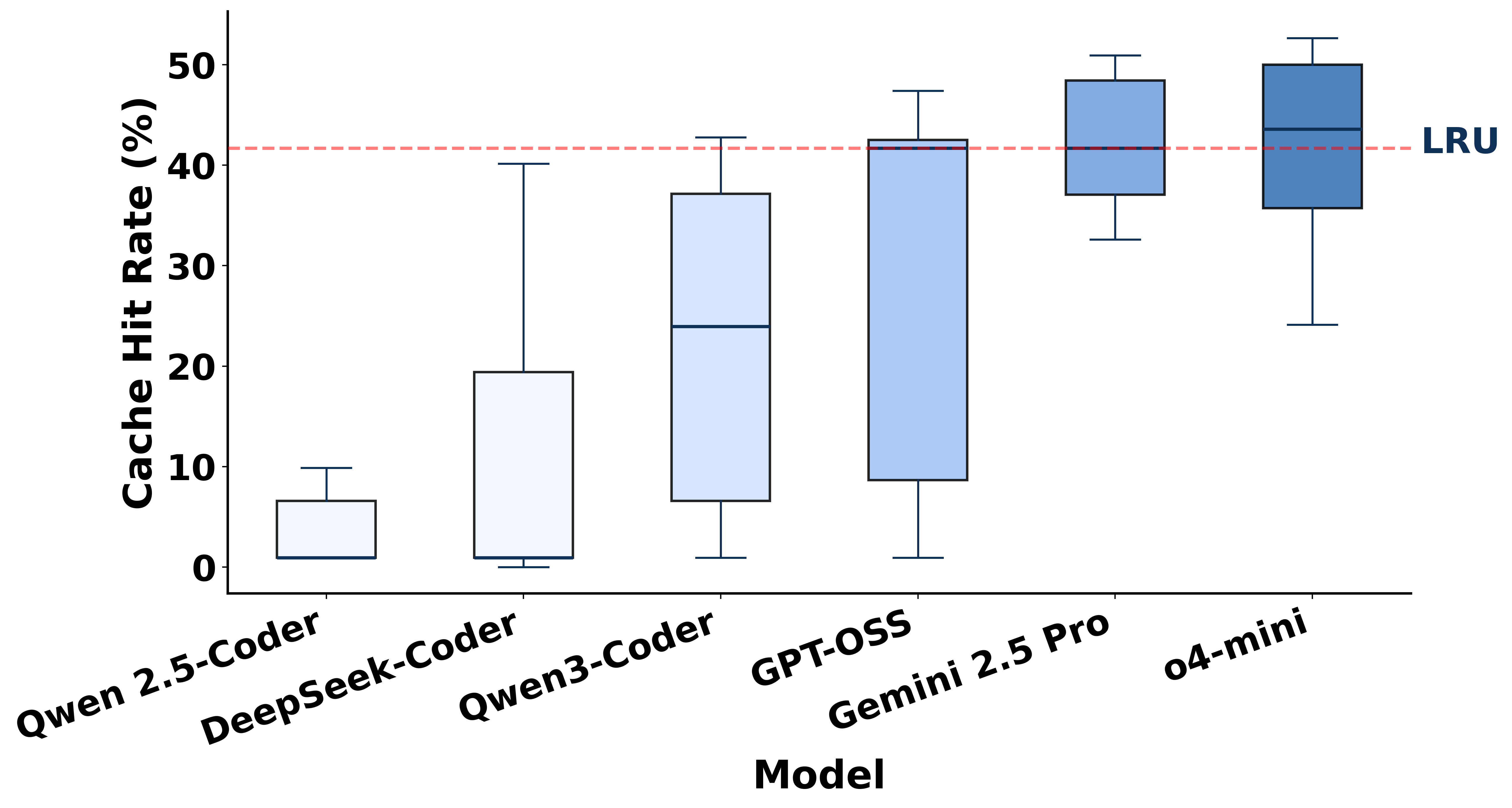}
\caption{Performance across various LLM models 
}
  \label{fig:models}
\end{figure}

\noindent\textbf{Impact of LLM Variations.} 
We evaluated \textsc{CacheForge}'s performance across multiple LLMs. Figure~\ref{fig:models} shows that \texttt{Gemini 2.5 Pro} and \texttt{o4-mini} yield the most consistent top-tier policies (48–52\% hit rate), highlighting the growing benefit of scaling and specialization in LLM-driven microarchitectural co-design.

\noindent\textbf{Memory, Exploration, \& Exploitation.}
\textsc{CacheForge} begins in exploration, where the LLM proposes novel
policy structures to surpass state-of-the-art baselines. With memory
enabled, prior policies and their embeddings are retrieved to guide this
search, helping the model reuse successful ideas and avoid redundant
paths. After a fixed number of unsuccessful trials, the system switches
to exploitation, refining the best policy found so far; here, lineage
memory captures its evolution and supports targeted improvements. Memory
has its strongest impact during exploration, enabling broader and more
non-redundant idea generation, while exploitation behaves similarly with
or without memory.

\begin{figure}[!htbp]
    \centering
    \includegraphics[width=\linewidth, height=3cm]{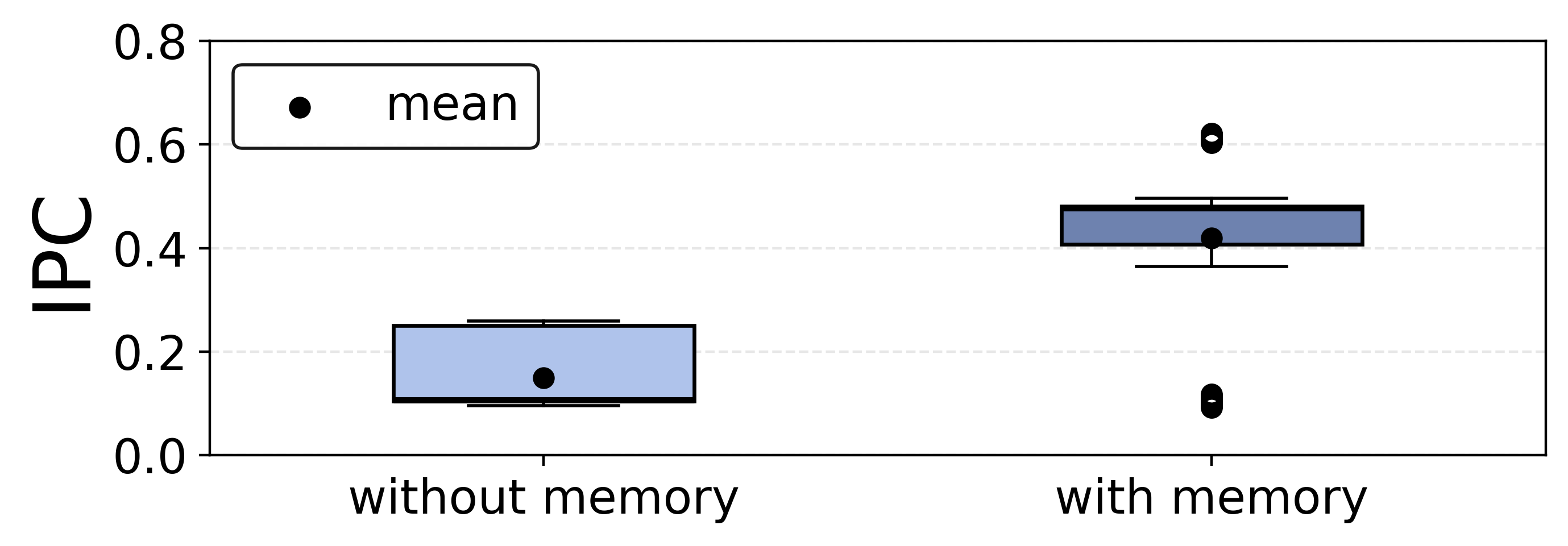}
    \caption{Impact of memory-augmented design.}
    \label{fig:memory}
\end{figure}
Figure~\ref{fig:memory} compares results with and without memory across
IPC, 
memory raises the mean
hit rate from 42.16\% to 48.27\% and increases mean IPC from
0.15 to 0.42. Speedup improves to 2.80$\times$, and even normalized by storage,
speedup-per-KB increases by 41.36\%. Overall, memory-guided exploration
produces higher-quality, more efficient, and more stable replacement
policies across the design space.

\input{Results/model}
\input{Results/temperate_onhit}

\begin{figure}[!htbp]
    \centering
    \includegraphics[width=1\linewidth, height=4.5cm]{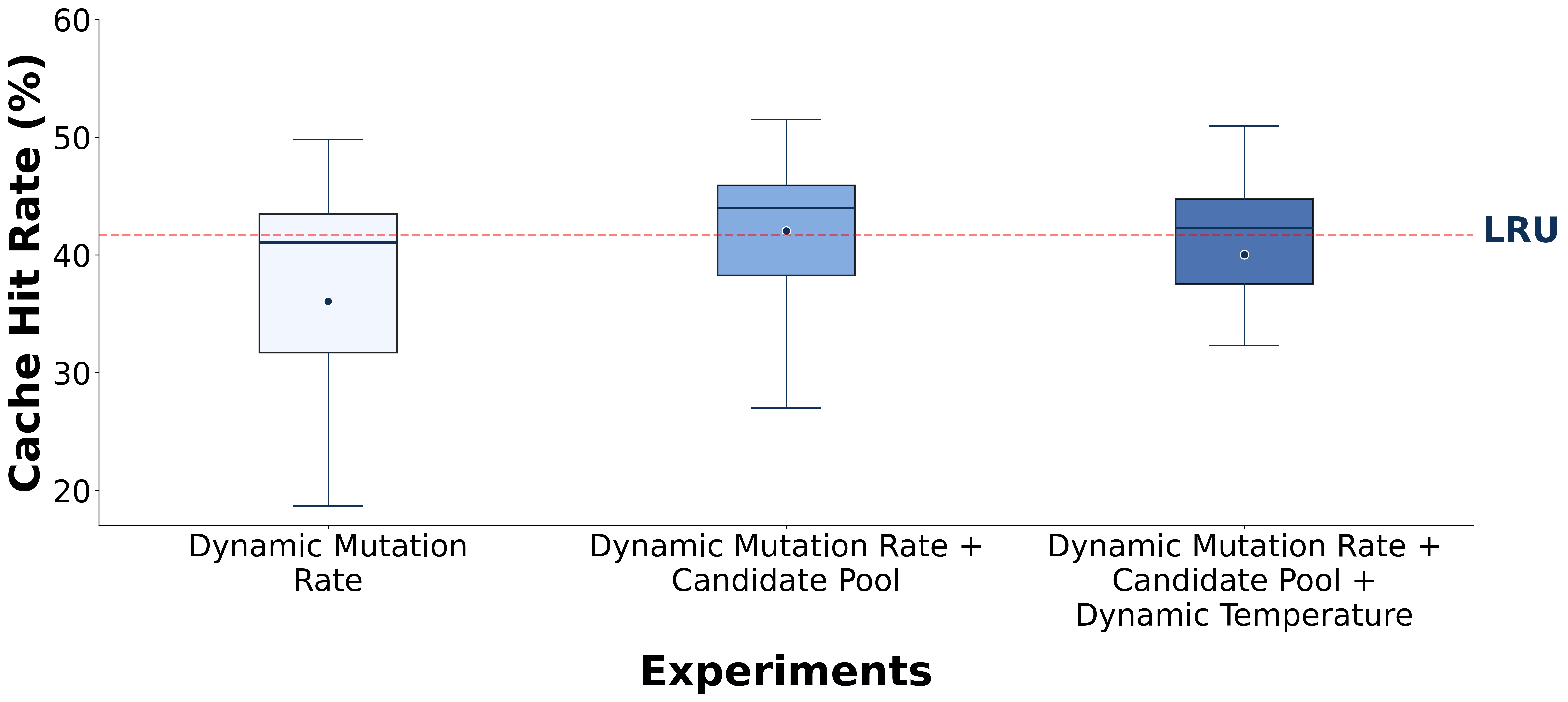}
    \caption{Impact of  dynamic mutation rate and temperature. }
    \label{fig:sensitivity_study}
\end{figure}

\noindent\textbf{Mutation Sensitivity Analysis.}
We performed a three-part ablation to assess how surrounding conditions
affect mutation-driven policy generation. Figure~\ref{fig:sensitivity_study}
compares hit-rate distributions for three mutation strategies. The
baseline dynamic mutation rate yields a median hit rate of 
41\% but
shows high variability, indicating unstable policy quality. Adding a
candidate pool raises the median to 
44\%, narrows the interquartile
range, and shifts the upper quartile upward, producing more consistent
and generally stronger policies. Incorporating dynamic temperature
alongside the candidate pool maintains similar variability but lifts the
minimum hit rate while slightly lowering the median to 
43\% and
reducing the concentration of top-end policies. Overall, the candidate
pool is the most effective enhancement, improving both median quality
and stability of generated policies.

\begin{figure}[!htbp]
    \centering
    \includegraphics[width=1\linewidth, height=3.5cm]{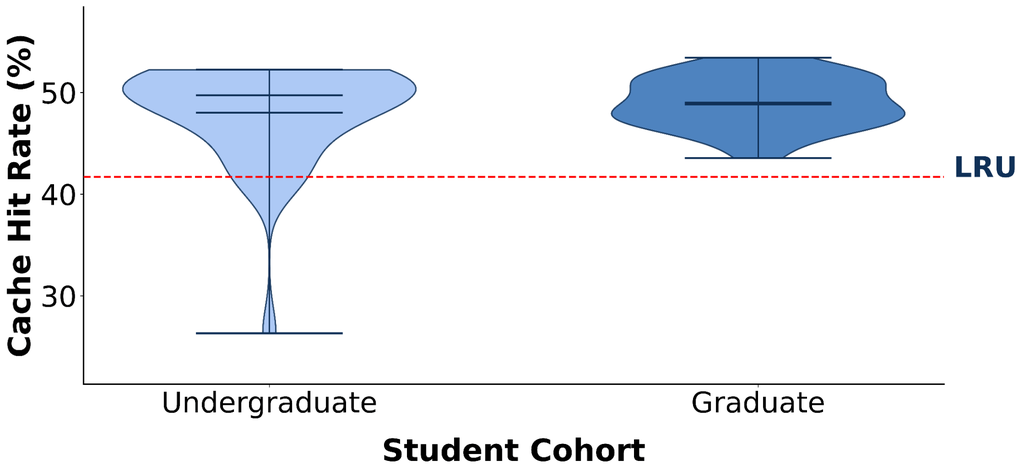}
    \caption{
Cache hit rate distributions for the undergraduate and graduate cohorts. and the LRU baseline.}
    \label{fig:student_violin_plot}
\end{figure}

\noindent\textbf{Impact of Prompt and Non-Expert Users.}
To evaluate whether \textsc{CacheForge} requires microarchitectural expertise to produce high-quality replacement policies, we conducted a user study with 60 students, 34 undergraduates and 26 graduate students, none of whom had prior exposure to cache design. Each participant interacted with the same LLM-guided workflow used throughout this paper, but was restricted to modifying only the \textit{prompt}. The goal of this experiment was to isolate how much of \textsc{CacheForge}’s performance depends on the skill of the human operator versus the autonomous generator–evaluator loop.
Undergraduates achieved a median hit rate of 49.75\% 
, with 88.2\% of participants generating policies strictly better than LRU. Graduate students performed even more uniformly, obtaining a median hit rate of 48.81\%
, with \emph{every} participant (26/26) surpassing LRU (Figure~\ref{fig:student_violin_plot}). 
These results demonstrate that \textsc{CacheForge}'s performance stems from its iterative generation-evaluation and causal feedback loop, eliminating the need for microarchitectural expertise. Furthermore, the framework provides significant educational value by illustrating how specific components impact replacement policies and CPU performance.



\vspace{-1em}

\section{Conclusion}
\label{sec:conclusion}
Prior work in cache optimization 
is limited by diminishing results, and is stuck within human-defined search spaces. 
This paper is the first to bridge this divide, using an LLM as a generative engine while simultaneously using a hardware simulator as a governance loop to ground the LLM's reasoning and enforce practical constraints such as hardware overhead.



\bibliographystyle{IEEEtranS}
\bibliography{references}

\end{document}